\documentclass[slac_one]{revtex4}
\usepackage{graphicx}
\usepackage{fancyhdr}
\def\mevc  {\ifmmode {\rm MeV}/c \else MeV$/c$\fi}
\def\mevcc {\ifmmode {\rm MeV}/c^2 \else MeV$/c^2$\fi}
\def\gevc  {\ifmmode {\rm GeV}/c \else GeV$/c$\fi}
\def\gevcc {\ifmmode {\rm GeV}/c^2 \else GeV$/c^2$\fi}
\def\ra    {\rightarrow}
\def\Dsm   {\ensuremath{D_s^-}}
\def\Bz    {\ensuremath{B^0}}
\def\Bzb   {\ensuremath{\bar{B}^0}}
\def\Bs    {\ensuremath{B_s^0}}
\def\Bsb   {\ensuremath{\bar{B}_s^0}}
\def\BsH   {\ensuremath{B_s^H}}
\def\BsL   {\ensuremath{B_s^L}}
\def\Bc    {\ensuremath{B_c^-}}
\def\Sb    {\ensuremath{\Sigma_b}}
\def\Xib   {\ensuremath{\Xi_b^-}}
\def\Omb   {\ensuremath{\Omega_b^-}}
\def\Lb    {\ensuremath{\Lambda_b^0}}
\def\dGoG  {\ensuremath{\Delta\Gamma_s/\Gamma_s}}
\def\dGs   {\ensuremath{\Delta\Gamma_s}}
\def\dMs   {\ensuremath{\Delta m_s}}
\def\BsJPsiPhi  {\ensuremath{\Bs\ra J/\psi\phi}}

\def\bsj   {\ensuremath{\beta_s^{J/\psi\phi}}}
\def\bssm  {\ensuremath{\beta_s^{SM}}}
\def\psnp  {\ensuremath{\phi_s^{NP}}}
\usepackage{relsize}
\def\babar{\mbox{\slshape B\kern-0.1em{\smaller A}\kern-0.1em
    B\kern-0.1em{\smaller A\kern-0.2em R}}}

\begin{document}

\title{Properties of Heavy \boldmath{$B$} Hadrons} 

\author{Manfred Paulini}
\affiliation{Carnegie Mellon University, Pittsburgh, PA 15213, USA}

\begin{abstract}
  We review recent measurements of heavy $B$~hadron states including
  masses and lifetimes of the \Bc~meson as well as excited
  $B$~states ($B^{**}$, $B_s^{**}$). We discuss properties of the
  \Bs~meson such as lifetime, lifetime difference \dGoG\ and
  $CP$~violation in \BsJPsiPhi~decays. We also summarize new
  measurements of the masses and lifetimes of bottom baryons including the
  \Lb~baryon, the \Sb~baryon states as well as the \Xib\ and \Omb~baryons.
\end{abstract}

\maketitle

\thispagestyle{fancy}


\section{INTRODUCTION} 

Hadrons containing bottom quarks can be classified according to their
$J^P$~quantum numbers. There are the ground state $0^-$~mesons such as
the neutral \Bzb~meson with quark content $|b\bar d\,\rangle$, the
charged $B^-$ (\,$|b\bar u\,\rangle$\,), 
the \Bsb~(\,$|b\bar s\,\rangle$\,) and the \Bc~meson which contains a
bottom and charm quark (\,$|b\bar c\,\rangle$\,).  
In addition, there are excited vector states
with spin-1 such as the $1^-$ states $\bar B^{*0}\ (\,|b\bar
d\,\rangle\,)$, $B^{*-}\ (\,|b\bar u\,\rangle\,)$, $\bar B_s^{*0}\
(\,|b\bar s\,\rangle\,)$, the $1^+$ states $\bar B^0_1\ (\,|b\bar
d\,\rangle\,)$, $\bar B_{s1}^{0}\ (\,|b\bar s\,\rangle\,)$ and the
$J^P=2^+$ states $\bar B^{*0}_2\ (\,|b\bar d\,\rangle\,)$, $\bar
B_{s2}^{*0}\ (\,|b\bar s\,\rangle\,)$.  Also, there exist bound $|b\bar
b\,\rangle$~mesons such as the $J^P=1^-$ states $\Upsilon(1S)$,
$\Upsilon(2S)$, $\Upsilon(3S)$ and the $\Upsilon(4S)$ resonance which is
the source of $\Bzb/B^-$~mesons at the KEKB and PEP-II
$e^+e^-$~$B$~factories with the Belle and \babar~experiments. The $0^-$
state $\eta_b$ $(\,|b\bar b\,\rangle\,)$, recently discovered at
\babar~\cite{Aubert:2008vj}, is discussed in detail in
Ref.~\cite{Grenier:2008gv} contributed to this conference.

In addition to $B$~mesons states, there exist baryons containing
$b$~quarks.  The lowest baryon state is the \Lb\ with quark content
$|bdu\,\rangle$.  Other bottom baryons with $J^P=1/2^+$ are the \Xib\
(\,$|bds\,\rangle$\,) as well as the $\Sb^-$ (\,$|bdd\,\rangle$\,) and
the $\Sb^+$ (\,$|buu\,\rangle$\,) plus their $3/2^+$ excited states
$\Sb^{*-}$ (\,$|bdd\,\rangle$\,) and $\Sb^{*+}$ (\,$|buu\,\rangle$\,).
In this paper, a heavy $B$~hadron is defined as all $B$~states
outlined above except for the \Bz\ and $B^-$~mesons. Since the
properties of $|b\bar b\,\rangle$ states are covered by other
presentations at this conference~\cite{Grenier:2008gv,Pakhlova:2008di},
we shall focus on \Bs, \Bc~mesons and excited $B$~states (generically
called $B^{**}$, $B_s^{**}$) as well as bottom baryons including the
\Lb~baryon, the \Sb~baryon states plus the \Xib\ and \Omb.

After defining ``heavy $B$~hadron'', we explain what is meant by
``properties'' of $B$~hadrons. Under properties we understand masses,
lifetimes and decay properties of heavy $B$~hadrons. This brings us to
the question of ``why study $B$~hadron states''?  A physicist typically
first comes into contact with the discussion of states while studying
the hydrogen atom in quantum mechanics. The spectroscopy of the H-atom
is explained as transitions between the various energy levels of the
hydrogen atom. This prime example of quantum mechanics allows us to draw
parallels to the study and spectroscopy of $B$~hadrons. The hydrogen
atom consists of a heavy nucleus in the form of the proton which is
surrounded by a light electron. The spectrum of the hydrogen atom
is sensitive to the interaction between proton and electron, which is
based on the 
electromagnetic Coulomb interaction and described by Quantum
Electrodynamics in its ultimate form. In analogy, a $B$~hadron consists
of a heavy bottom quark surrounded either by a light anti-quark, to form
a $B$~meson or a di-quark pair, to form a bottom baryon. The interaction
between the $b$~quark and the other quark(s) in a $B$~hadron is based on
the strong interaction or Quantum Chromodynamics (QCD). It is often
stated that heavy quark hadrons are the hydrogen atom of QCD. The study
of $B$~hadron states is thus the study of (non-perturbative) QCD,
providing sensitive tests of potential models, heavy quark effective
theory (HQET) and all aspects of QCD, including lattice gauge
calculations.

\subsection{\boldmath{$B$} Hadron Lifetimes}

In the spectator model of $B$~hadron decay, the $b$ quark decays like a
free particle. The other (anti-)quark(s) in the hadron act as pure
spectators without influencing the $b$~quark decay. In such a simple
weak decay picture, the lifetimes of all $B$~hadrons would be equal. In
reality, the strong force in the form of gluons, coupling to the quarks,
as well as final state interactions, influence the pure weak decay.
Measurements of $B$~hadron lifetimes thus study the interplay between
the strong and weak interaction. Predictions of $B$~hadron lifetimes are
provided in the heavy quark expansion (HQE) which in turn allows us to
expand the inclusive decay width $\Gamma_B$ in powers $1/m_b$ of the
bottom quark mass
\begin{equation}
\Gamma_{B} \sim |V_{CKM}|^2 \sum_n c_n(\mu)\left(\frac{1}{m_b}\right)^n\, 
\langle H_b|O_n|H_b\,\rangle.
\end{equation}
In HQE short distance effects contained in the Wilson coefficients
$c_n(\mu)$, evaluated in perturbation theory, are separated from long
distance physics represented by the matrix element $\langle
H_b|O_n|H_b\,\rangle$ to be computed through non-perturbative QCD sum
rules, operator product expansion methods or lattice QCD calculations.
In HQE the order ${\cal O}(1/m_b^2)$ distinguishes meson versus baryon
decays while spectator effects of order ${\cal O}(1/m_b^3)$
differentiate between the lifetimes of \Bz, $B^+$ and \Bs~mesons. These
calculations allow for precise predictions of $B$~hadron lifetimes where
many can be found in the literature. Reference~\cite{blifepred} only
quotes a few of them. Most of these predictions can be summarized in
form of the following estimates for $B$~hadron lifetime ratios
\begin{equation}
\frac{\tau(B^+)}{\tau(\Bz)}=1.06\pm0.02, \quad\quad\quad
\frac{\tau(\Bs)}{\tau(\Bz)}=1.00\pm0.01, \quad\quad\quad
\frac{\tau(\Lb)}{\tau(\Bz)}=0.88\pm0.05.
\end{equation}
Measurements of $B$~hadron lifetimes thus test the validity of HQE, a
technique which is also used to supply input for the extraction of
elements of the Cabibbo-Kobayashi-Maskawa (CKM) quark mixing matrix.

\begin{figure*}[tb]
\centering
\includegraphics[height=71mm]{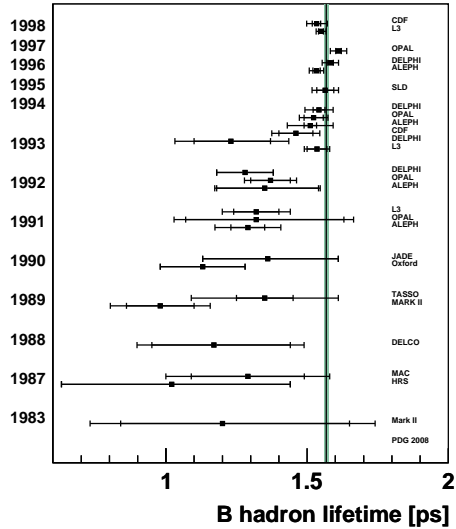}
\caption{History of measurements of the average $B$~hadron lifetime.}
\label{fig:BlifeHistory}
\end{figure*}

Since history always provides guidance, Figure~\ref{fig:BlifeHistory}
shows the history of measurements of the average $B$~hadron lifetime
which starts with the first measurement of the average lifetime of
bottom hadrons~\cite{Lockyer:1983ev} in 1983. The Mark\,II detector
measured $\tau_b=(1.20^{+0.45}_{-0.36}\pm0.30)$~ps which is within large
errors in agreement with the current average $B$~hadron lifetime as
determined by the Particle Data Group (PDG)~\cite{ref:PDG2008}. However,
Figure~\ref{fig:BlifeHistory} indicates that all early measurements of
bottom hadron lifetimes appear to obtain low central values compared to
the current world average until the availability of precision
measurements pined down the current world average. Such an effect seems
to repeat itself in other $B$~hadron lifetime measurements as we shall
see later.

\section{Experimental Environment
\label{sec:tev}}

The producers of hadrons containing $b$~quarks are currently the KEKB
and PEP-II $e^+e^-$ colliders together with the Belle and
\babar~experiments, as well as the Fermilab Tevatron where the CDF and
D0~experiments are operating. At the Fermilab Tevatron all $B$~hadrons
are produced.  Besides the \Bz\ and $B^+$~meson, which are the only
bottom hadrons produced at the $B$~factories operating at the
$\Upsilon(4S)$ resonance, the Tevatron is a source for \Bs\ and
$B_c^+$~mesons as well as baryons containing $b$~quarks such as the \Lb,
\Sb\ or \Xib. It has been common believe that the study of
\Bs~properties is the domain of the CDF and D0 experiments operating at
the Tevatron.

However, it has recently become possible to produce \Bs~events in
sufficiently large numbers in $e^+e^-$~collisions at the $\Upsilon(5S)$
resonance which can decay into pairs of $\Bs\Bsb$, $B_s^{*0}\Bsb$ or
$B_s^{*0}\bar{B}_s^{*0}$. The KEKB collider operating at the
$\Upsilon(5S)$ resonance at a center-of-mass energy of $\sim\!10.87$~GeV
has delivered a dataset to the Belle detector in 2005 and 2006 totaling
an integrated luminosity of $(23.6\pm0.3)$~fb$^{-1}$. From a
study~\cite{belle:2008sc} of $161\pm15$ reconstructed decays
$\Bs\ra\Dsm\pi^+$, Belle has reported the measurement of the branching
fraction ${\cal B}(\Bs\ra \Dsm\pi^+)=[3.67^{+0.35}_{-0.33}\,{\rm
  (stat.)}^{+0.43}_{-0.42}\,{\rm (syst.)}\pm0.49\,(f_s)]\times10^{-3}$,
where the largest systematic error, which is due to the uncertainty in
the production fraction $f_s=N_{B_s^{(*)}\bar{B}_s^{(*)}}/N_{b\bar b}$,
is quoted separately.  The obtained branching fraction is compatible
with the CDF result~\cite{ref:PDG2008,cdf:2006qw} and is slightly higher
than ${\cal B}(\Bz\ra D^-\pi^+)$ by $1.3\,\sigma$. In 
addition, Belle observes $6.7^{+3.4}_{-2.7}$ signal events from decays
$\Bs\ra D_s^{\mp}K^{\pm}$ and measures the branching ratio 
${\cal B}(\Bs\ra D_s^{\mp}K^{\pm})=[2.4^{+1.2}_{-1.0}\,{\rm (stat.)}
\pm0.3\,{\rm (syst.)}\pm0.3\,(f_s)]\times10^{-4}$ with a significance of
$3.5\,\sigma$.

After a successful 1992-1996 Run\,I data taking period of the Fermilab
Tevatron (for a review of $B$~physics results from e.g.~CDF in Run\,I
see Ref.~\cite{Paulini:1999px}), the Tevatron operates in Run\,II at a
centre-of-mass energy of 1.96~TeV with a bunch crossing time of 396~ns
generated by $36\times36$ $p\bar p$ bunches. The initial Tevatron
luminosity steadily increased from 2002 to 2008 with a peak luminosity
of $>30\cdot 10^{31}$~cm$^{-2}$s$^{-1}$ reached in 2008.  The total
integrated luminosity delivered by the Tevatron to CDF and D0 at the
time of this conference is $\sim\!4.5$~fb$^{-1}$ with about
$3.7$~fb$^{-1}$ recorded to tape by each collider experiment.  However,
most results presented in this review use about 1-3~fb$^{-1}$ of data.
The features of the CDF and D0 detectors are described elsewhere in
References~\cite{Acosta:2004yw} and \cite{Abazov:2005pn},
respectively. For the remainder of this paper, we will focus on results
from the Tevatron experiments CDF and D0.

\section{PROPERTIES OF \boldmath{\Bs}~MESONS} 

In the neutral \Bs~system there exist two flavour eigenstates, the 
$\Bs=|\bar b s\,\rangle$ and $\Bsb=|b\bar s\,\rangle$. The time
evolution of these states is governed by the Schr\"odinger equation
\begin{equation}
i\frac{d}{dt}
\left(\begin{array}{c}
|\Bs(t)\,\rangle \\ |\Bsb(t)\,\rangle \\ 
\end{array}\right)
=\left[{\bf M}-\frac{i}{2}\,{\bf \Gamma}\,\right]
\left(\begin{array}{c}
|\Bs(t)\,\rangle \\ |\Bsb(t)\,\rangle \\ 
\end{array}\right)
\quad\quad
{\rm{with}}\ \ 
{\bf{M}}=
\left(\begin{array}{cc}
M_0 & M_{12} \\ M_{12}^* & M_0 \\
\end{array}\right)
\quad
{\rm{and}}\ \ 
\bf{\Gamma}=
\left(\begin{array}{cc}
\Gamma_0 & \Gamma_{12} \\ \Gamma_{12}^* & \Gamma_0 \\
\end{array}\right),
\end{equation}
where {\bf M} is the mass matrix and $\bf \Gamma$ is the decay matrix.
The mass eigenstates \BsH\ and \BsL\ are admixtures of the flavour
eigenstates \Bs\ and \Bsb:
\begin{equation}
|\BsH\,\rangle = p\,|\Bs\,\rangle - q\,|\Bsb\,\rangle,\quad\quad
|\BsL\,\rangle = p\,|\Bs\,\rangle + q\,|\Bsb\,\rangle,\quad {\rm{with}}\ \
\frac{q}{p}=\frac{V_{tb}^*V_{ts}}{V_{tb}V_{ts}^*}.
\end{equation}
The fact that the mass eigenstates are not the same as the flavour
states gives rise to oscillations between the \Bs\ and \Bsb~states with a
frequency proportional to the mass difference of the mass eigenstates,
$\dMs=m_H-m_L\sim2\,|M_{12}|$. In the standard model (SM)
particle-antiparticle oscillations are explained in terms of
second-order weak processes involving virtual massive particles that
provide a transition amplitude between the \Bs\ and \Bsb~states. The
decay width difference between the mass eigenstates
$\dGs=\Gamma_L-\Gamma_H\sim2\,|\Gamma_{12}|\cos\phi_s$ is related to the
$CP$~phase $\phi_s=\rm{arg}(-M_{12}/\Gamma_{12})$.  Assuming no
$CP$~violation in the \Bs~system, which is justified in the standard
model where the $CP$~phase is expected to be small ($\phi_s^{\rm
  SM}\sim0.004$~\cite{Lenz:2006hd}), the \Bs~mass eigenstates are also
$CP$~eigenstates where $\Gamma_L$ is the width of the $CP$~even state
corresponding to the short lived state in analogy to the kaon system
where the short-lived state ($K^0_S$) is $CP$~even.  $\Gamma_H$ is the
width of the $CP$~odd state corresponding to the long lived \Bs~state.

Thus the experimental observables describing the \Bs~system are the
masses $m_H$ and $m_L$ of the \Bs~mass eigenstates accessible through a
measurement of the mass difference \dMs\ in \Bs-\Bsb~oscillations. Other
experimental quantities are the width difference \dGs, the total
decay width $\Gamma_s=(\Gamma_H+\Gamma_L)/2=1/\tau_s$, which is related
to the mean \Bs~lifetime $\tau_s$, as well as the $CP$~phase~$\phi_s$.

\subsection{Measurements of  the \boldmath{\Bs}~Meson Lifetime} 

In the light of a substantial width difference \dGs, the \Bs~system
contains short- and long-lived components similar to the kaon system and
various \Bs~decay channels will have different proportions of the
\BsH\ and \BsL~eigenstates. Lifetime measurements of different
final states have therefore different meaning and can be broken down
into several categories. First, there are flavour specific decays, such as
semileptonic $\Bs\ra\Dsm\ell^+\nu$  or $\Bs\ra\Dsm\pi^+$ decays, which
have equal fractions of \BsL\ and \BsH\ at proper time zero from where
both components will evolve with their specific lifetimes
$\tau_H=1/\Gamma_H$ and  $\tau_L=1/\Gamma_L$. Fitting a single
exponential to such a decay distribution measures the flavour specific
lifetime 
\begin{equation}
\tau(\Bs)_{\rm flav.spec.} = \frac{1}{\Gamma_s}
\frac{1+\left(\frac{\dGs}{2\Gamma_s}\right)^2}
{1-\left(\frac{\dGs}{2\Gamma_s}\right)^2}.
\end{equation}
Second, there is the $CP$~specific lifetime measured in decays that are
assumed to be either $CP$~even or $CP$~odd. For example, the exclusive
decay $\Bs\ra K^+K^-$ is expected to be $CP$~even within 5\% and
measures the lifetime of the light mass eigenstate
$\tau_L=1/\Gamma_L$. In 2006, CDF reported a preliminary measurement of
$\tau(\Bs)=(1.53\pm0.18\pm0.02)$~ps from $\Bs\ra K^+K^-$. Finally,
there are decays into a mixed $CP$~final state where it is possible to
disentangle the final state $CP$~components. For example, an angular
analysis can be used to decompose the $CP$~components in the exclusive
decay \BsJPsiPhi\ which is expected to be dominated by the $CP$~even
state and its lifetime.

\subsubsection{\Bs~Flavour Specific Lifetime} 
\label{sec:bsflav}

\begin{figure*}[tb]
\centering
\includegraphics[height=71mm]{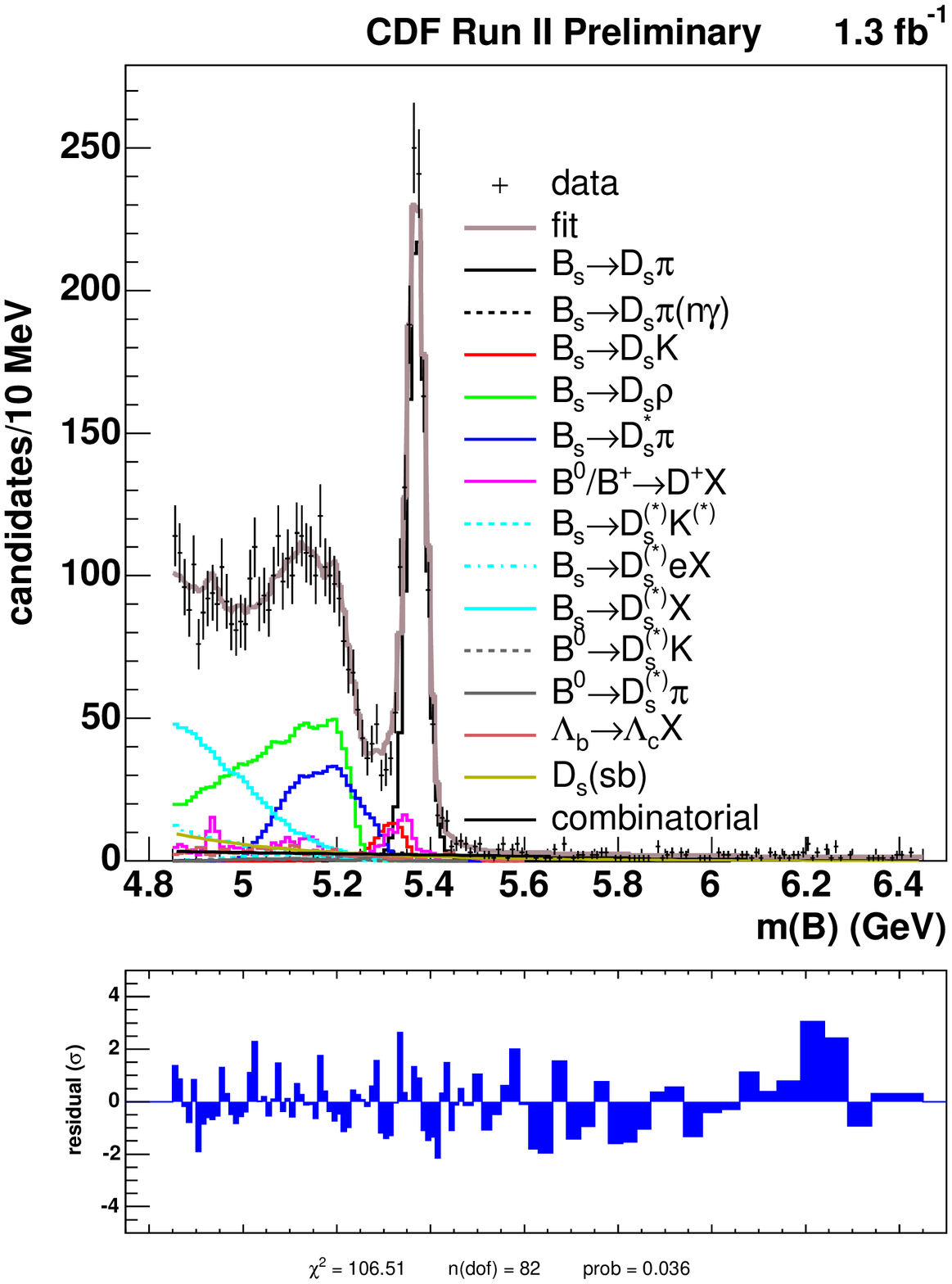}
\includegraphics[height=71mm]{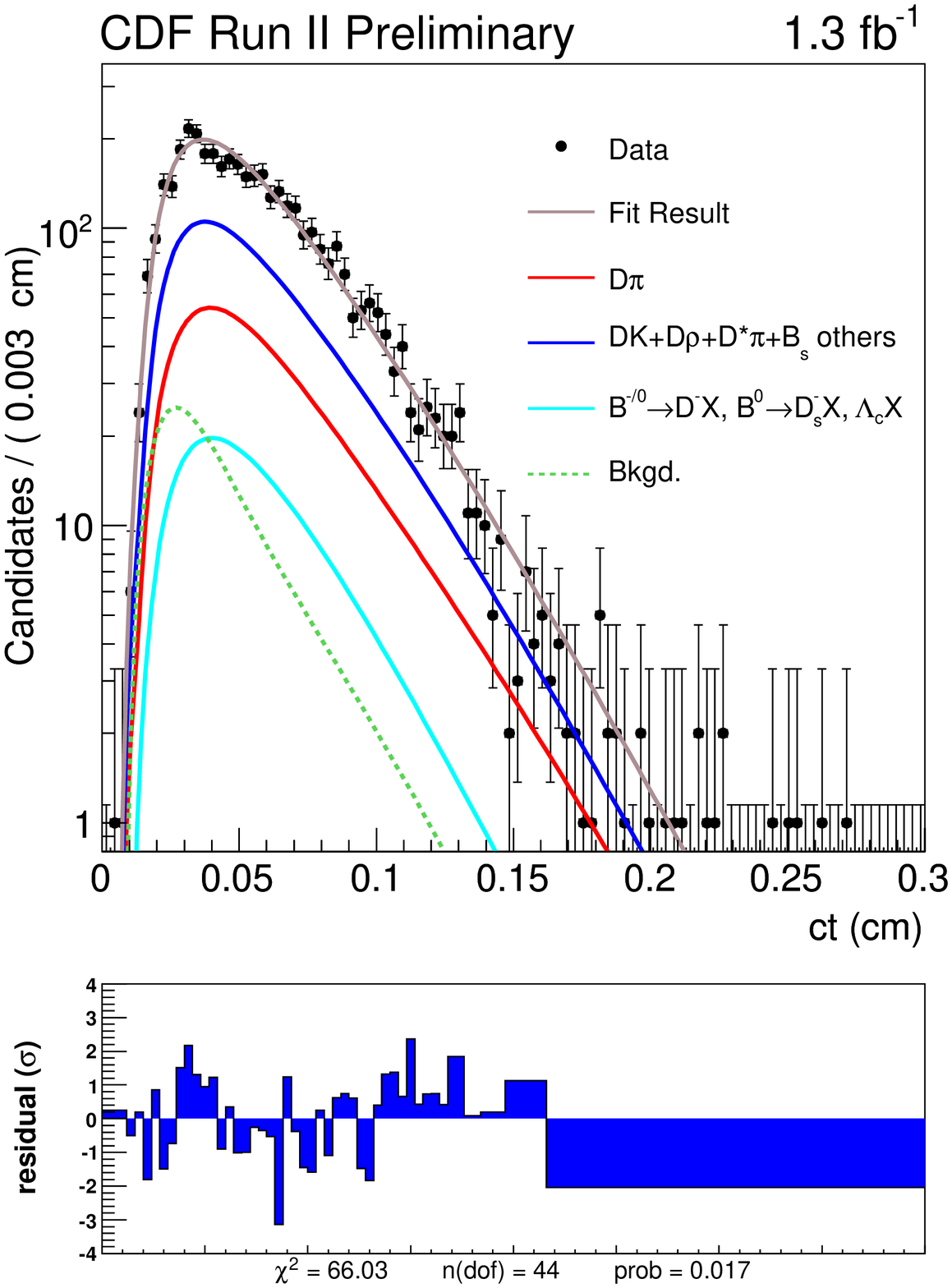}\ \ \ \ \ \
\includegraphics[height=70mm]{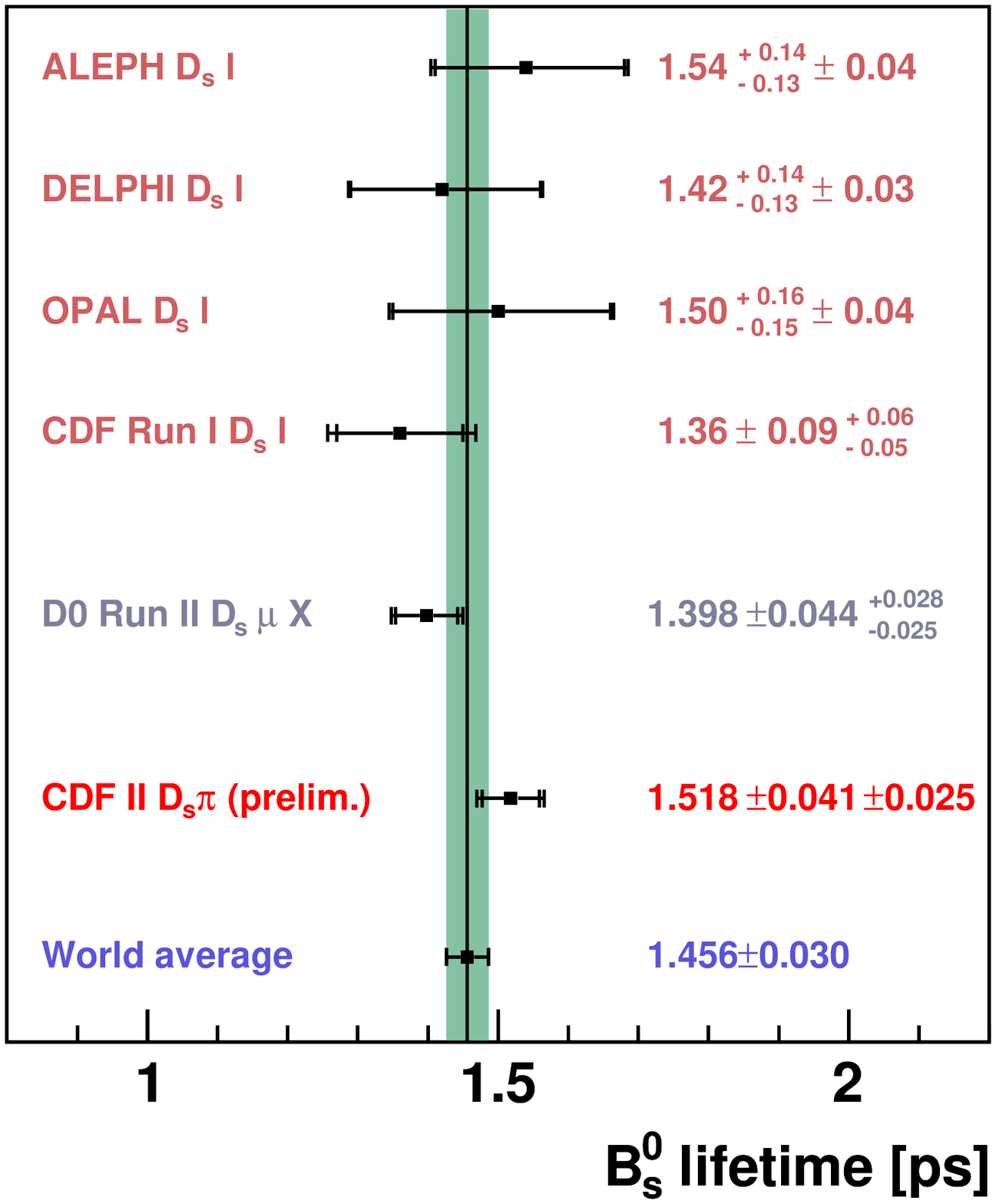}
\put(-462,177){\large\bf (a)}
\put(-215,177){\large\bf (b)}
\put(-120,183){\large\bf (c)}
\caption{(a) Mass fit of events reconstructed as
  $\Bs\ra\Dsm(\phi\pi^-)\pi^+$ in the CDF measurement of the flavour
  specific \Bs~lifetime and (b) projection of the corresponding lifetime
  fit results. (c) Summary of \Bs~flavour specific lifetime measurements.}
\label{fig:Bslife}
\end{figure*}

CDF presented a new measurement of the \Bs~flavour specific lifetime at
this conference. In a data sample of $\sim\!1.3$~fb$^{-1}$ of $p\bar p$
collisions, CDF reconstructs \Bs~candidates through its decay
$\Bs\ra\Dsm\pi^+$ followed by $\Dsm\ra\phi\pi^-$ with $\phi\ra K^+K^-$. 
This sample yields more than 1100 fully reconstructed \Bs~candidates as
shown in Fig.~\ref{fig:Bslife}(a). In addition, this sample also
includes about 2000 partially reconstructed \Bs~candidates that
contribute to the fit of the lifetime distribution shown in
Fig.~\ref{fig:Bslife}(b). CDF obtains the flavour specific lifetime
$\tau(\Bs)=(1.518\pm0.041\pm0.027)$~ps. The ratio of this
result and the world average \Bz~lifetime~\cite{ref:PDG2008} yields
$\tau(\Bs)/\tau(\Bz)=0.99\pm0.03$ in good agreement with theoretical
predictions. Note, using the 2008 PDG mean
\Bs~lifetime~\cite{ref:PDG2008} results in
$\tau(\Bs)/\tau(\Bz)=0.95\pm0.02$. A compilation of all \Bs~flavour
specific lifetime measurements to date is given in
Fig.~\ref{fig:Bslife}(c) and the world averaged flavour specific
\Bs~lifetime including the new CDF result is determined to be
$\tau(\Bs)=(1.456\pm0.030)$~ps~\cite{Barberio:2008fa}.

\subsubsection{\Bs~Lifetime from \BsJPsiPhi\ and Measurement of \dGs} 

The decay \BsJPsiPhi\ is the transition of the spin-0 pseudo-scalar \Bs\
into two spin-1 vector particles. The orbital angular momenta of the
vector mesons, $J/\psi$ and $\phi$, can be used to distinguish the
$CP$~even $S$-wave ($L=0$) and $D$-wave ($L=2$) final states from the
$CP$~odd $P$-wave ($L=1$) final state. Typically the set of decay angles
$\vec{\rho}=(\cos\theta_T,\phi_T,\cos\psi_T)$ defined in the
transversity basis (see e.g.~Ref.~\cite{Aaltonen:2007gf}) is used to
disentangle the $CP$~mixture of the $J/\psi\phi$ final state. Such an
angular decomposition reveals that the decay is dominated by the
$CP$~even state.
 
\begin{figure*}[tb]
\centering
\includegraphics[height=77mm]{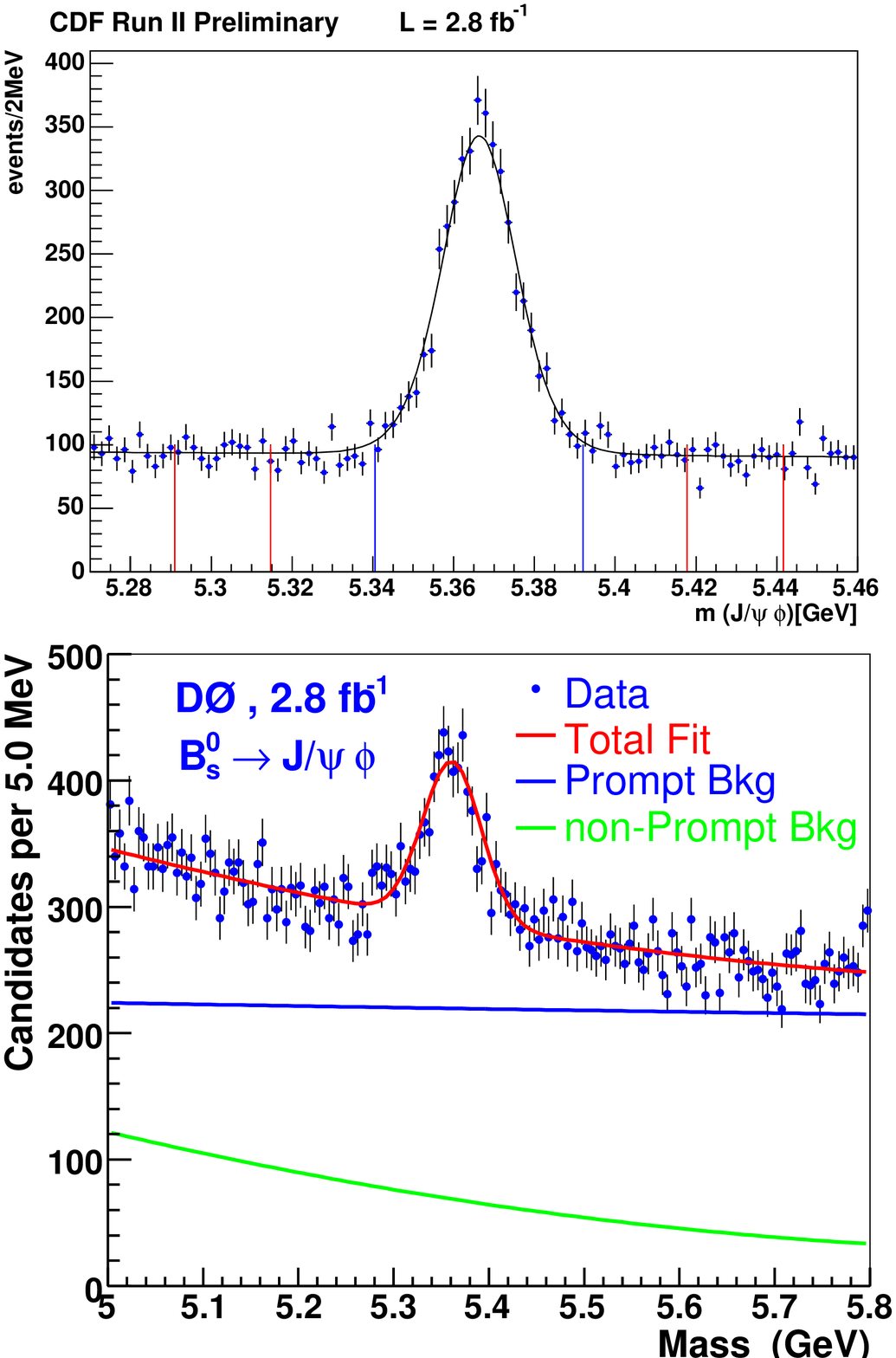}
\includegraphics[height=77mm]{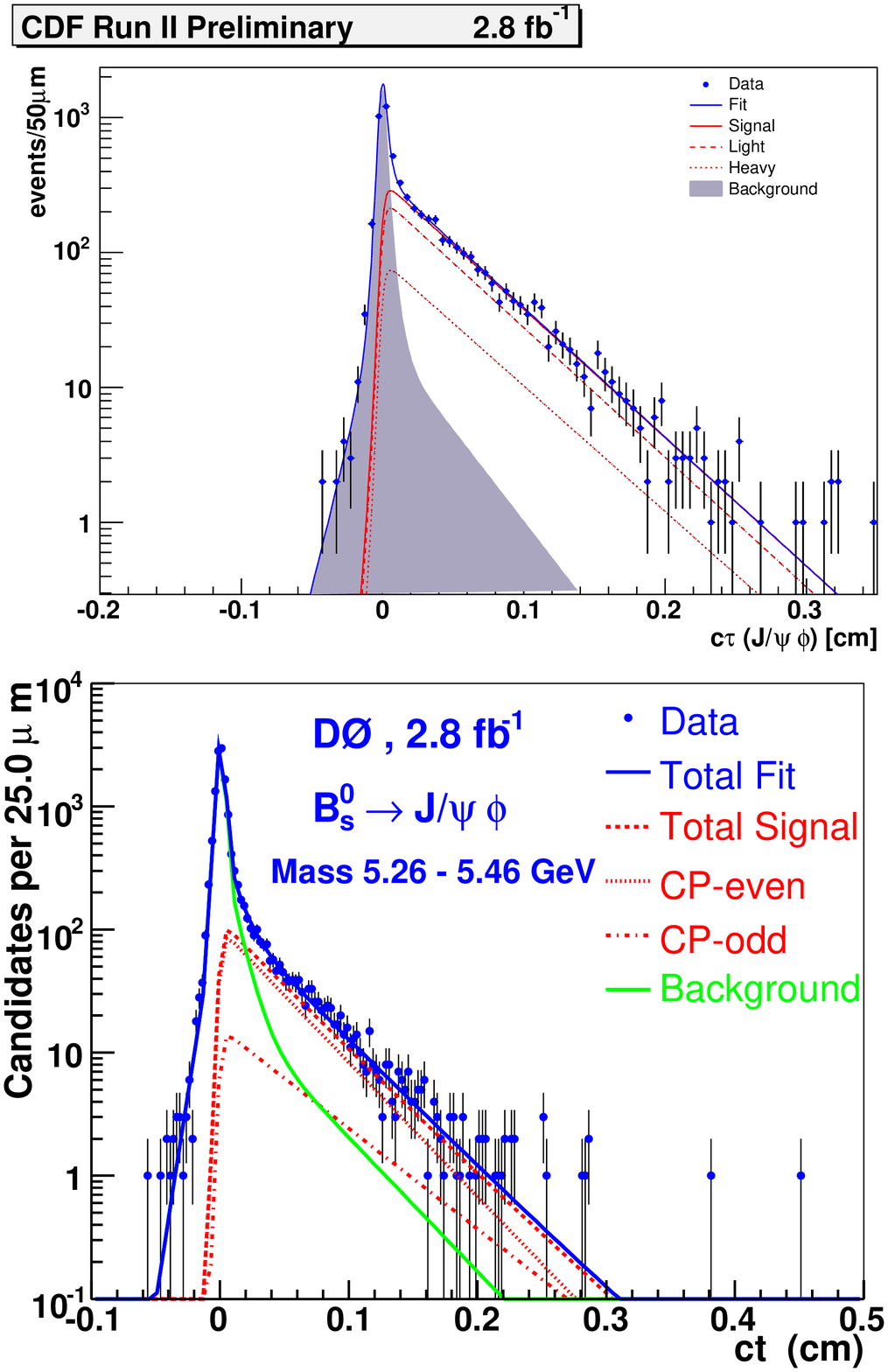}\ \ \ \
\includegraphics[height=77mm]{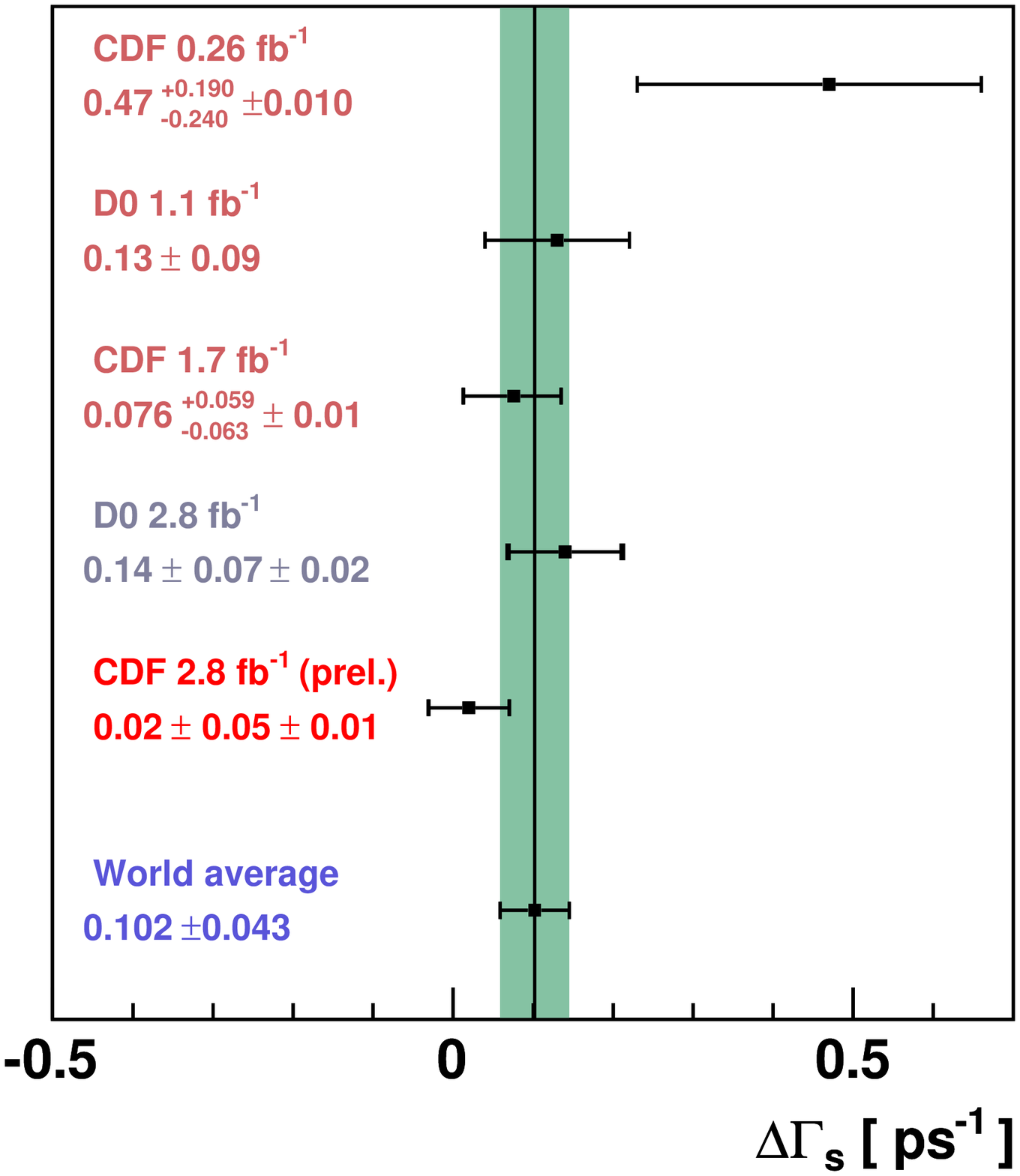}
\put(-470,197){\large\bf (a)}
\put(-320,195){\large\bf (b)}
\put(-30,180){\large\bf (c)}
\caption{(a) Invariant mass distribution of \BsJPsiPhi\ from CDF (top)
  and D0 (bottom). (b) Corresponding lifetime distributions with fit
  projections overlaid from CDF (top) and D0 (bottom). Note the
  difference in the large $CP$~even and the much smaller $CP$~odd
  contributions. (c) Compilation of various \dGs~measurements.}
\label{fig:BsdG}
\end{figure*}

The experimental situation with respect to measurements of the mean
\Bs~lifetime $\tau_s=2/(\Gamma_H+\Gamma_L)$ from \BsJPsiPhi\ assuming no
$CP$~violation is as follows: The D0~collaboration has
published~\cite{abazov:2008fj} a result based on 2.8~fb$^{-1}$ of data,
while the CDF collaboration updated their published
result~\cite{Aaltonen:2007gf} based on 1.35~fb$^{-1}$ for this
conference with a preliminary result using 2.8~fb$^{-1}$ of data. The
D0~analysis identifies $1967\pm65$~$J/\psi\phi$ signal
events~\cite{abazov:2008fj} as shown in Figure~\ref{fig:BsdG}(a) while
CDF finds $3166\pm56$ \Bs~signal events in 2.8~fb$^{-1}$ of data.  With
these events D0 measures a mean \Bs~lifetime
$\tau_s=(1.53\pm0.06\pm0.01)$~ps and quotes
$\dGs=(0.14\pm0.07\pm0.02)$~ps$^{-1}$ assuming no $CP$~violation in the
\Bs~decay. The corresponding numbers from the preliminary CDF analysis
are $\tau_s=(1.53\pm0.04\pm0.01)$~ps and
$\dGs=(0.02\pm0.05\pm0.01)$~ps$^{-1}$. As can be seen in the lifetime
distributions of Fig.~\ref{fig:BsdG}(b), the lifetime distribution is
mainly $CP$~even while the $CP$~odd component is much smaller. A
compilation of various measurement of \dGs\ is shown in
Figure~\ref{fig:BsdG}(c). The preliminary CDF result mentioned above 
is not yet included in the world average of
$\dGs=(0.102\pm0.043)$~ps$^{-1}$~\cite{Barberio:2008fa}.  When these
direct measurements of \dGs\ are combined with the \Bs~flavour specific
lifetime measurements discussed in Sec.~\ref{sec:bsflav}, a constrained
result of $\dGs=(0.067^{+0.031}_{-0.035})$~ps$^{-1}$ is
obtained~\cite{Barberio:2008fa}.

\subsection{\boldmath{$CP$}~Violation in \boldmath{\BsJPsiPhi}} 

In analogy to measurements of the time dependent $CP$~asymmetry in
neutral \Bz~decays into e.g.~$\Bz\ra J/\psi K_S^0$ accessing the $CP$
violating phase $\sin(2\beta)$ which arises through the interference
between decay and mixing, the application of flavour tagging to
\BsJPsiPhi\ events measures the corresponding phase in \Bs~decays.  This
phase, which is responsible for $CP$~violation in \BsJPsiPhi\ in the
standard model, is in analogy to the phase $\sin(2\beta)$ called
$\sin(2\bssm)$ and is defined as 
$\bssm={\rm  arg}(-V_{ts}V_{tb}^*/V_{cs}V_{cb}^*)$.  In the context of
the standard model, this phase is expected to be small and global fits
of experimental data constrain it to
$2\bssm\sim0.04$~\cite{Lenz:2006hd,Barberio:2008fa}. Measuring such a
small value of $\sin(2\bssm)$ is currently beyond the experimental reach
at the Tevatron.  However, new physics may contribute significantly
larger values to the $CP$~violating phase in
\BsJPsiPhi~decays~\cite{Lenz:2006hd,Ligeti:2006pm,Hou:2006mx}.  In this
case, the observed $CP$~phase would be modified by a phase \psnp\ due to
new physics processes, and can be expressed as $2\bsj=2\bssm-\psnp$. If
$\psnp\gg2\bssm \Rightarrow \psnp\gg\phi_s$, we expect
$\dGs=2\,|\Gamma_{12}|\cos\phi_s\sim 2\,|\Gamma_{12}|\cos(2\bsj)$. We 
can then make the approximation for the observed quantities
$2\bsj=-\phi_s^{J/\psi\phi}$. The current interest in measuring
$CP$~violation in \BsJPsiPhi\ is therefore a search for enhanced
$CP$~violation through new physics processes.

\begin{figure*}[tb]
\centering
\includegraphics[height=60mm]{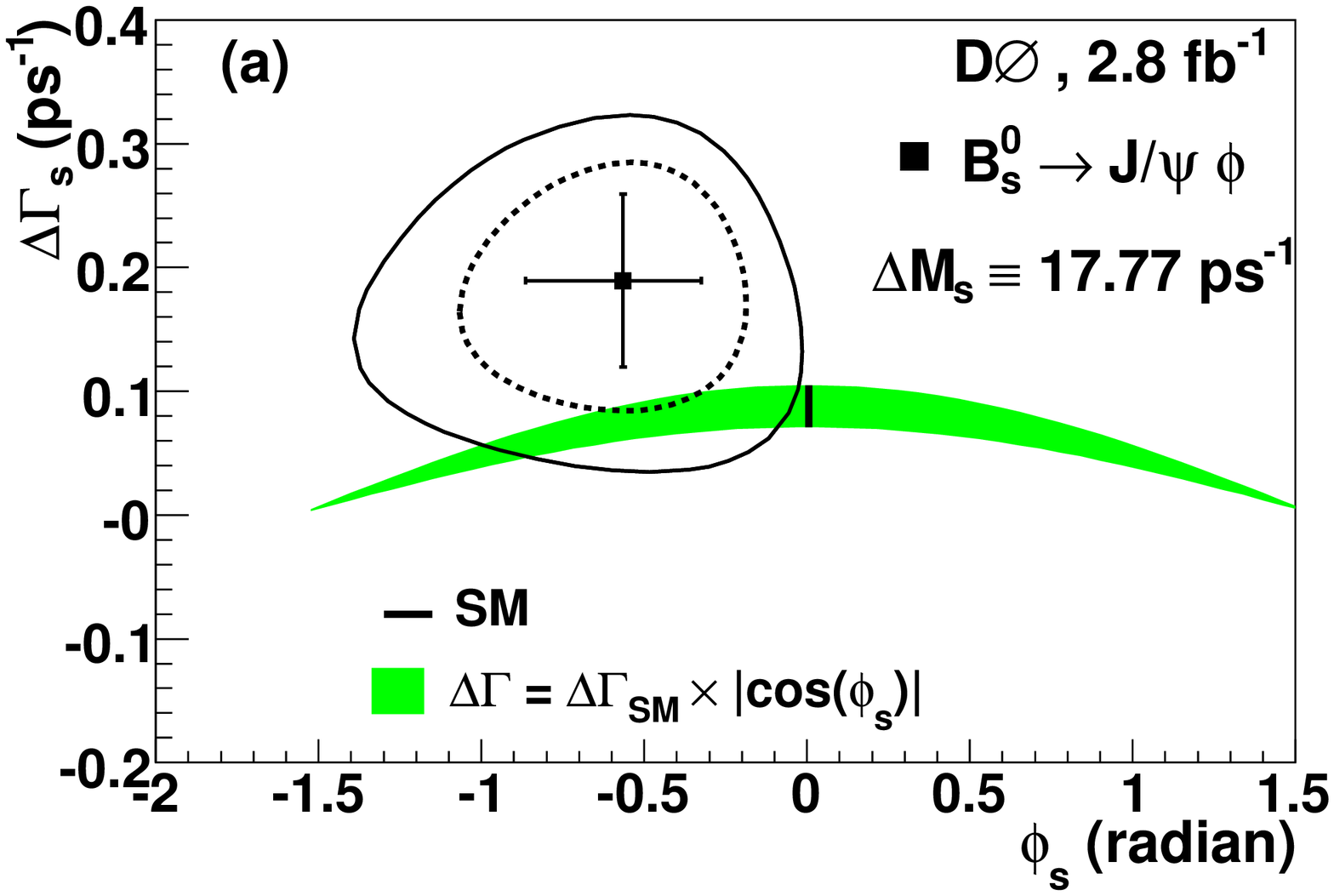}
\includegraphics[height=64mm]{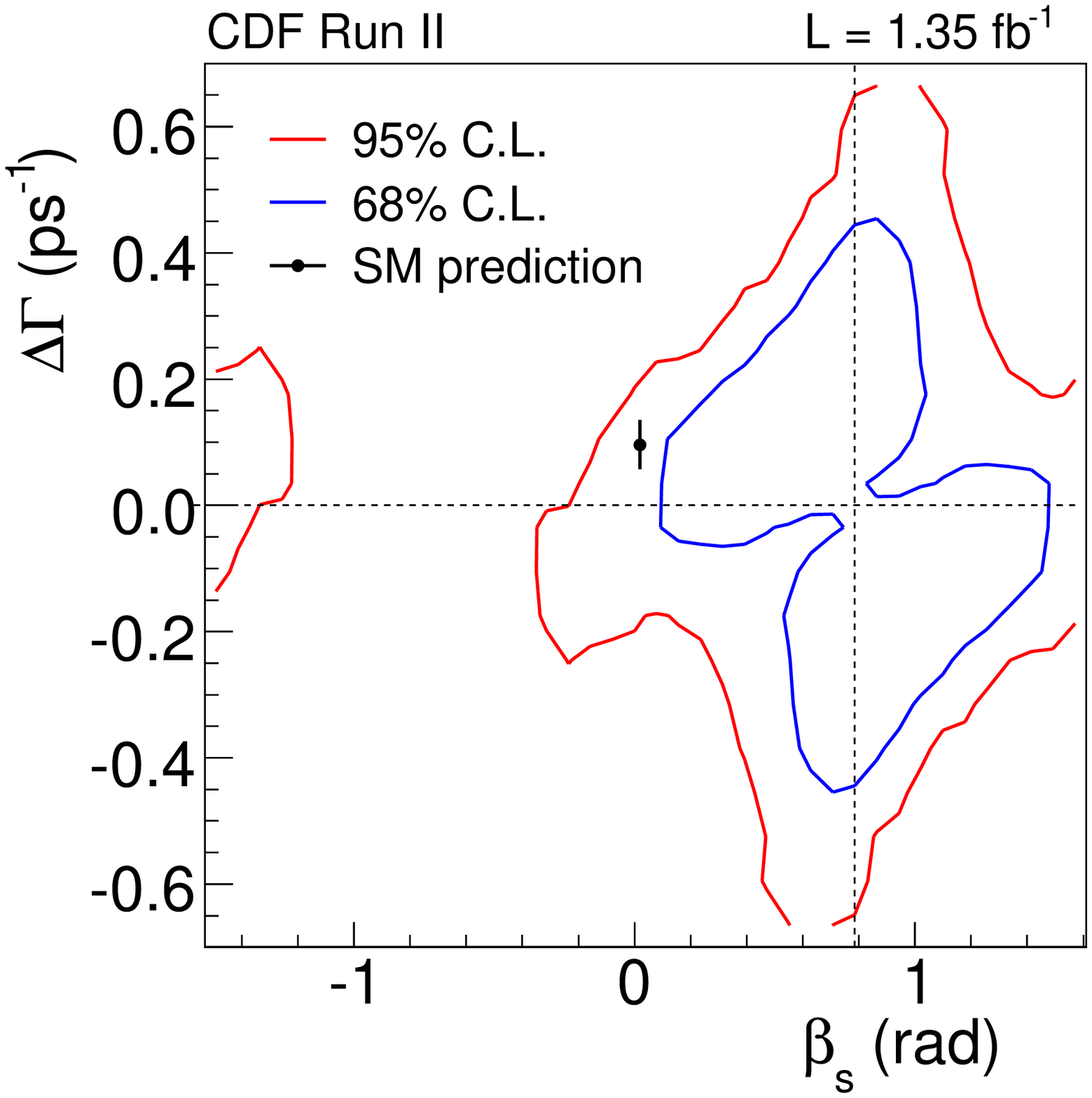}
\put(-90,150){\large\bf (b)}
\caption{Confidence level regions in plane of \dGs\ versus \bsj\ in
  flavour-tagged \BsJPsiPhi\ analysis from (a) the 2.8~fb$^{-1}$
  result from D0 and (b) the 1.35~fb$^{-1}$ result from CDF. Note the
  transformation $2\bsj=-\phi_s^{J/\psi\phi}$.}
\label{fig:Bscontour}
\end{figure*}

At the 2008 winter conferences both Tevatron experiments presented
tagged, time dependent angular analyses of \BsJPsiPhi~decays.  Due to
the non-parabolic behaviour of the log-likelihood function, no
meaningful point estimates for \bsj\ can be quoted and both experiments
construct their results as confidence level regions in the plane of
\dGs\ versus \bsj. The D0 result~\cite{abazov:2008fj} based on
2.8~fb$^{-1}$ of data is shown in Figure~\ref{fig:Bscontour}(a) while
the CDF result from 1.35~fb$^{-1}$ of data~\cite{Aaltonen:2007he} is
displayed in Figure~\ref{fig:Bscontour}(b). Both experiments observe a
mild inconsistency with the SM prediction
$2\bssm\sim0.04$. Interestingly, the CDF and D0 inconsistencies with the
standard model both point in the same direction. Assuming the SM
prediction, CDF quotes a probability of 15\% to observe a likelihood
ratio equal or higher than the one observed in data which corresponds to
about $1.5\,\sigma$. Using constraints on the strong phases, D0 finds a
$p$-value of 6.6\% corresponding to a $1.8\,\sigma$ inconsistency with
the SM hypothesis~\cite{abazov:2008fj}.

\begin{figure*}[tb]
\centering
\includegraphics[height=56mm]{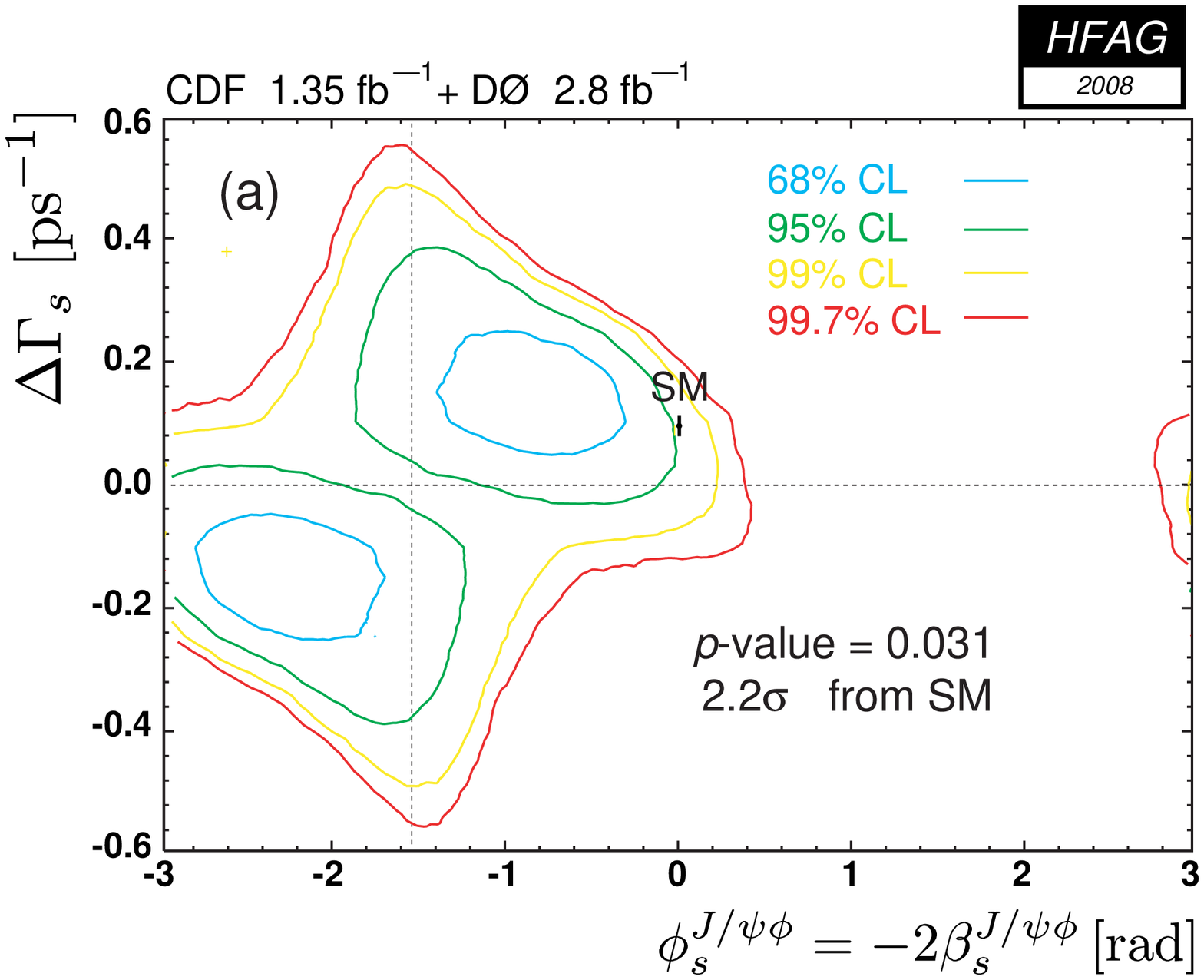}
\includegraphics[height=54mm]{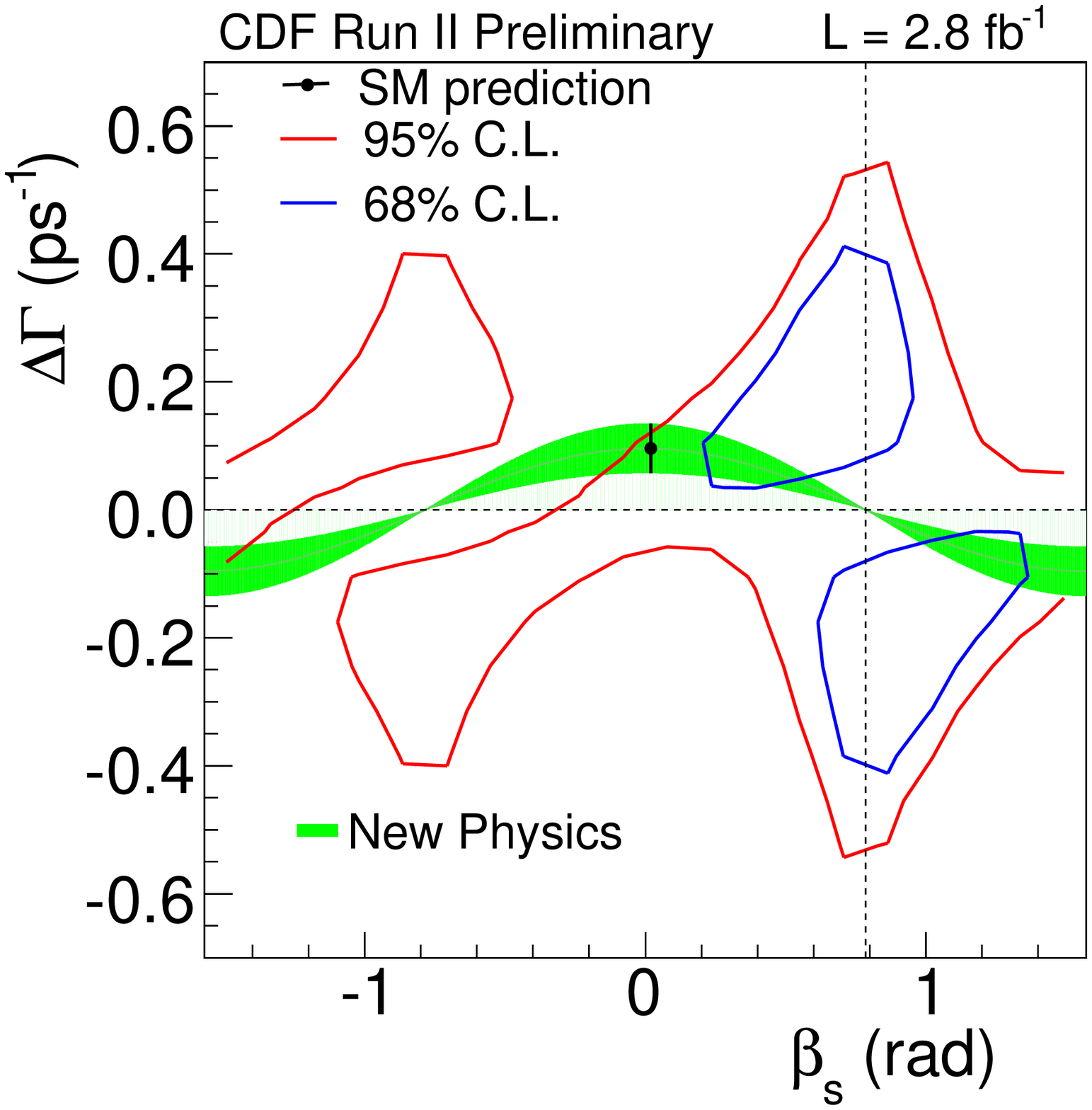}
\includegraphics[height=54mm]{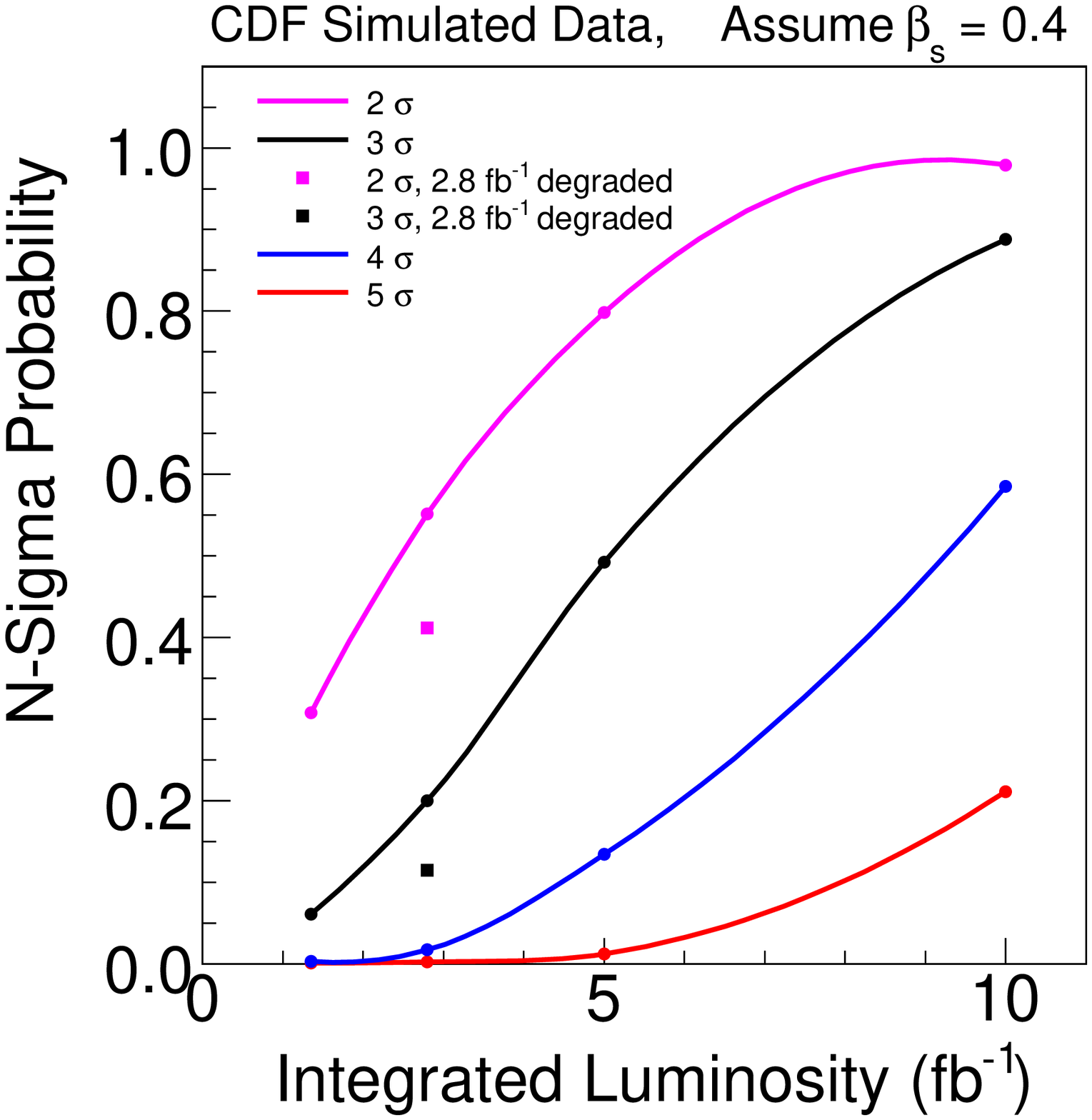}
\put(-475,125){\large\bf (a)}
\put(-220,125){\large\bf (b)}
\put(-115,95){\large\bf (c)}
\caption{Confidence regions in the plane of \dGs\ versus \bsj\ from (a)
  combination of CDF and D0 likelihoods obtained from their
  flavour-tagged \BsJPsiPhi\ analyses and (b) the CDF preliminary update
  of the \BsJPsiPhi~analysis based on 2.8~fb$^{-1}$. (c) CDF expected
  sensitivity to $\bsj=0.40$ for various integrated luminosities from
  1.35~fb$^{-1}$ to 10~fb$^{-1}$ of data.}
\label{fig:Bscombo}
\end{figure*}

There are two new results on $CP$~violation in \BsJPsiPhi\ presented at
this conference. First, D0 
released their data without a constraint on the strong phases allowing
for a combination of the CDF and D0 likelihoods obtained in their
flavour-tagged \BsJPsiPhi\ time-dependent analyses. The combined result
is shown in Figure~\ref{fig:Bscombo}(a) and restricts \bsj\ to the
interval $[0.14,0.73]\cup[0.83,1.42]$ at 90\% confidence level (CL). The
consistency of the combined result gives a
$p$-value of 3.1\% corresponding to a $2.2\,\sigma$ discrepancy
with the SM prediction. Second, CDF released an update of their
published analysis~\cite{Aaltonen:2007he} using 2.8~fb$^{-1}$ of
data. The new result again shown as a confidence region in the plane of
\dGs\ versus \bsj, as displayed in Figure~\ref{fig:Bscombo}(b),
confirms the trend of the published result. CDF finds
that the $p$-value at the SM expectation is 7\% corresponding to a
$\sim\!1.8\,\sigma$ discrepancy with the standard model. Furthermore,
CDF determines that the projected one-dimensional range for \bsj\ is
confined to the interval $\bsj \in [0.28,\,1.29]$ at 68\% CL. 

The updated CDF analysis was obtained with a suboptimal dataset which
did not allow for the use of particle identification for the entire data
selection and restricted the application of same side kaon flavour
tagging to the first half of the data.  To estimate the future reach of
this analysis, CDF's expected sensitivity of measuring a value of
$\bsj=0.40$ for various integrated luminosities ranging from
1.35~fb$^{-1}$ up to 10~fb$^{-1}$ of data is shown in
Figure~\ref{fig:Bscombo}(c). With about 5~fb$^{-1}$ of data, the
probability to measure an assumed value of $\bsj=0.40$ is about 50\%
which puts some excitement on awaiting further updates of the
measurement of \bsj\ in \BsJPsiPhi\ from the Tevatron.

\section{PROPERTIES OF OTHER HEAVY \boldmath{$B$}~MESONS} 

\subsection{Orbitally Excited \boldmath{$B$}~Mesons}

Until a couple of years ago, excited meson states containing $b$~quarks
had not been studied well. Only the stable $J^P=0^-$ ground states
$B^+$, \Bz\ and \Bs\ and the excited $1^-$ state $B^*$ had been firmly
established. Quark models predict the existence of two wide ($B_0^*$ and
$B_1^\prime$) and two narrow ($B_1^0$ and $B_2^{*0}$) bound
$P$-states~\cite{ref:Eichten}. The wide states decay via an $S$-wave and
therefore have a large width of a couple of hundred \mevcc, which makes
it difficult to distinguish such states from combinatoric
background. The narrow states decay via a $D$-wave transition ($L=2$)
and thus should have a small width of
$\sim\!10~\mevcc$~\cite{ref:Ebert,ref:Isguretal}. Almost all previous
observations~\cite{ref:BdsLEP_OPAL,ref:BdsLEP} of the narrow states
$B_1^0$ and $B_2^{*0}$ have been made indirectly using inclusive or
semi-exclusive $B$~decays which prevented the separation of both states
and a precise measurement of their properties. In contrast, the masses,
widths and decay branching fractions of these states are predicted with
good precision by theoretical models~\cite{ref:Ebert,ref:Isguretal}.

$B_1^0$ and $B_2^{*0}$ candidates are reconstructed in the following
decay modes: $B_1^0 \ra B^{*+}\pi^-$ with $B^{*+}\ra B^+\gamma$ and
$B_2^{*0} \ra B^{*+}\pi^-$ with $B^{*+}\ra B^+\gamma$ as well as
$B_2^{*0} \ra B^{+}\pi^-$. In both cases the soft photon from the $B^{*+}$
decay is not observed resulting in a shift of about 46~\mevcc\ in the
mass spectrum.  D0 reconstructs $B^+$ candidates in the fully
reconstructed mode $B^+\ra J/\psi K^+$ with $J/\psi\ra\mu^+\mu^-$, while
CDF selects $B^+$ mesons in addition through the $B^+\ra \bar{D}^0\pi^+$ and
$\bar{D}^0 \pi^+\pi^+\pi^-$~mode with $\bar{D}^0\ra K^+\pi^-$. The CDF
analysis~\cite{ref:CDF_orbB} is based on 1.7~fb$^{-1}$ of data resulting
in a $B^+\ra J/\psi K^+$ signal of $\sim\!51\,500$ events as well as $40\,100$
and 11\,000 candidates in the $\bar{D}^0\pi^+$ and $\bar{D}^0 \pi^+\pi^+\pi^-$
channels, respectively. The D0 measurement~\cite{ref:D0_orbB} employs
1.3~fb$^{-1}$ of Run\,II data and finds a signal peak of $23\,287\pm344$
events attributed to the decay $B^+\ra J/\psi K^+$.

\begin{figure*}[tb]
\centering
\includegraphics[height=54mm]{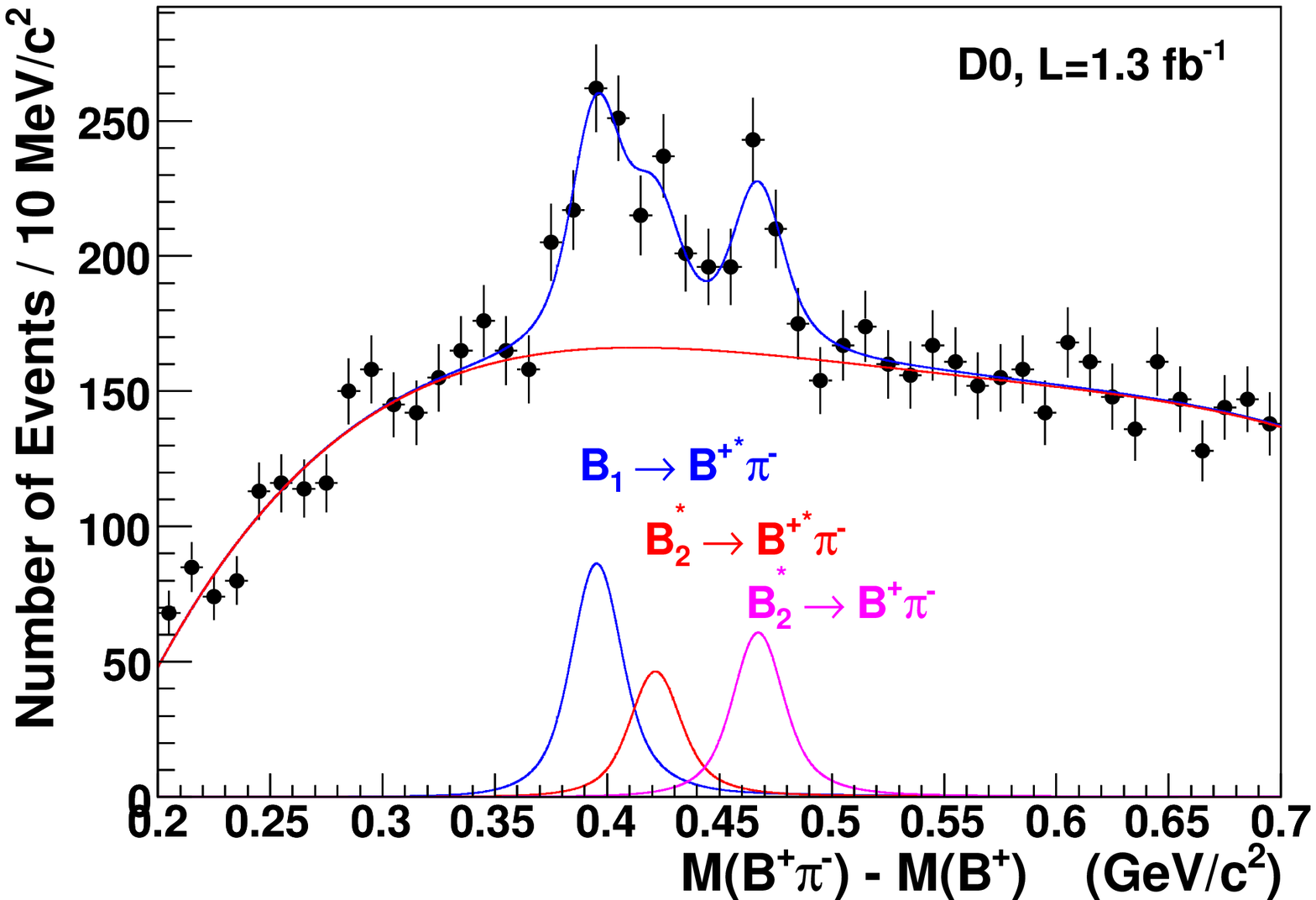}
\includegraphics[height=55mm]{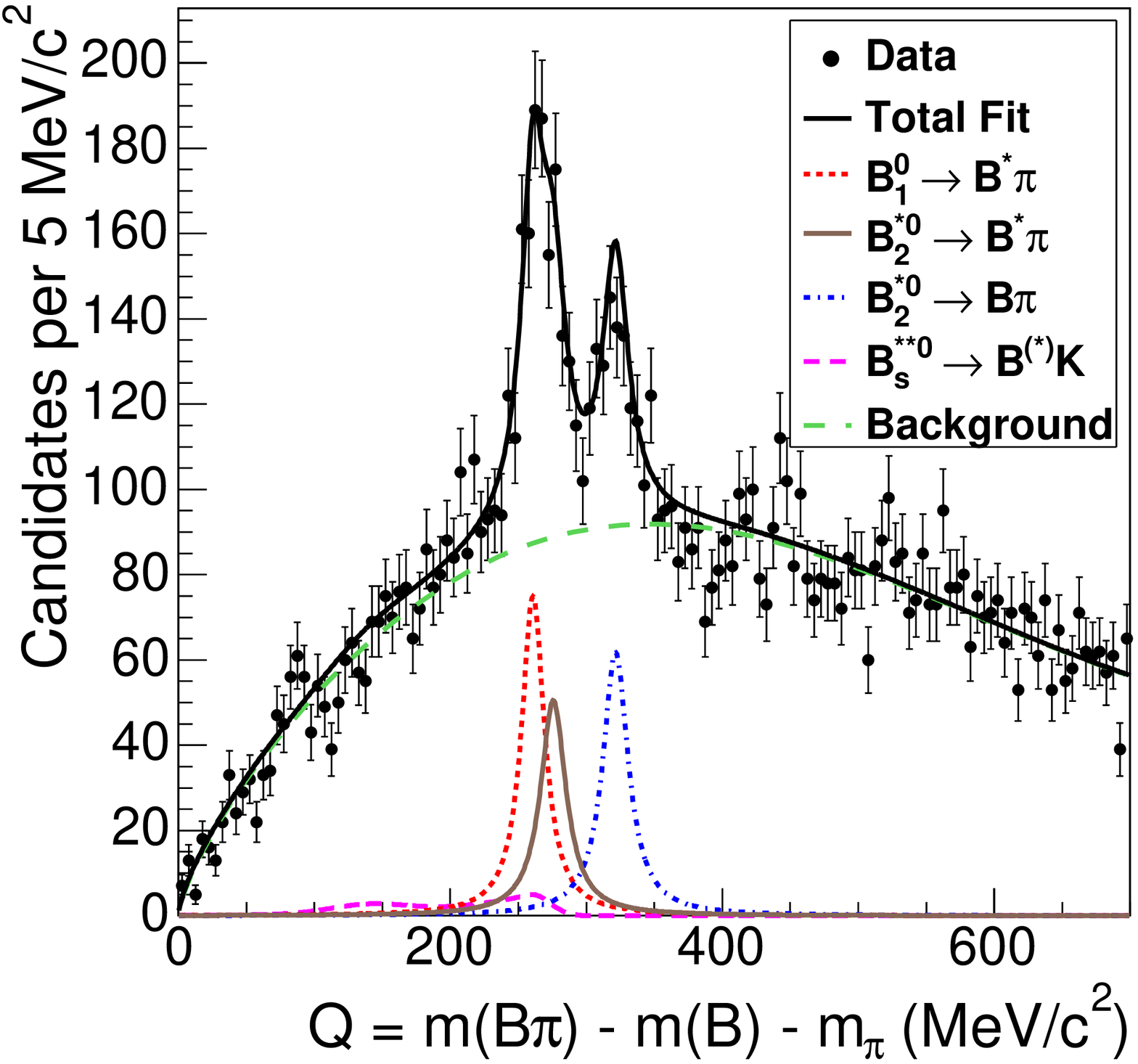}
\put(-360,135){\large\bf (a)}
\put(-130,138){\large\bf (b)}
\caption{Fit to the $B^{**}$ mass difference (a) $\Delta m =
  m(B\pi)-m(B)$ from D0 and (b) $Q = m(B\pi)-m(B)-m(\pi)$ from CDF.}
\label{fig:Bd_double_star}
\end{figure*}

D0 presents their measured mass distribution as $\Delta m =
m(B\pi)-m(B)$ as shown in Figure~\ref{fig:Bd_double_star}(a), while CDF
plots $Q = m(B\pi)-m(B)-m(\pi)$ as displayed in
Fig.~\ref{fig:Bd_double_star}(b). Clear signals for the narrow excited
$B^{**}$~states are observed: CDF reconstructs a total of about 1250~$B^{**}$
candidates while D0 observes a total of $662\pm91\pm140$ candidates for
the narrow $B^{**}$ states. The measured masses are reported as
$m(B_1^0)=(5720.6\pm2.4\pm1.4)$~\mevcc\ and
$m(B_2^{*0})=(5746.8\pm2.4\pm1.7)$~\mevcc\ from D0, while CDF quotes
$m(B_1^0)=(5725.3{^{+1.6}_{-2.2}}{^{+1.4}_{-1.5}})$~\mevcc\ and
$m(B_2^{*0})=(5740.2{^{+1.7}_{-1.8}}{^{+0.9}_{-0.8}})$~\mevcc.  Both
results are in modest agreement.

\subsection{Orbitally Excited Strange \boldmath{$B$}~Mesons}

The properties of $|b\bar s\,\rangle$ excited meson states, referred to
as $B_{s}^{**}$, and the comparison with the properties of excited
states in the $|b\bar u\,\rangle$ and $|b\bar{d}\,\rangle$ systems
provides good tests of various models of quark bound
states~\cite{ref:Eichten,ref:Ebert,ref:Falk95}. These models predict the
existence of two wide resonances ($B_{s0}^*$ and $B_{s1}^\prime$) and
two narrow ($B_{s1}^0$ and $B_{s2}^{*0}$) bound $P$-states. The wide
states decay through an $S$-wave and thus have a large width of order
hundred \mevcc. This makes it difficult to distinguish such states from
combinatoric background. The narrow states decay through a $D$-wave
($L=2$) and therefore should have a small width of order
1~\mevcc~\cite{ref:Isguretal} varying with predicted mass. If the mass
of the orbitally excited $B_{s}^{**}$ is large enough, then the main
decay channel should be through $B^{(*)}K$ as the $\Bs\pi$ decay mode is
not allowed by isospin conservation. Previous
observations~\cite{ref:BdsLEP_OPAL} of the narrow $B_{s}^{**}$
$P$-states have been made indirectly preventing the separation of both
states.

\begin{figure*}[tb]
\centering
\includegraphics[height=55mm]{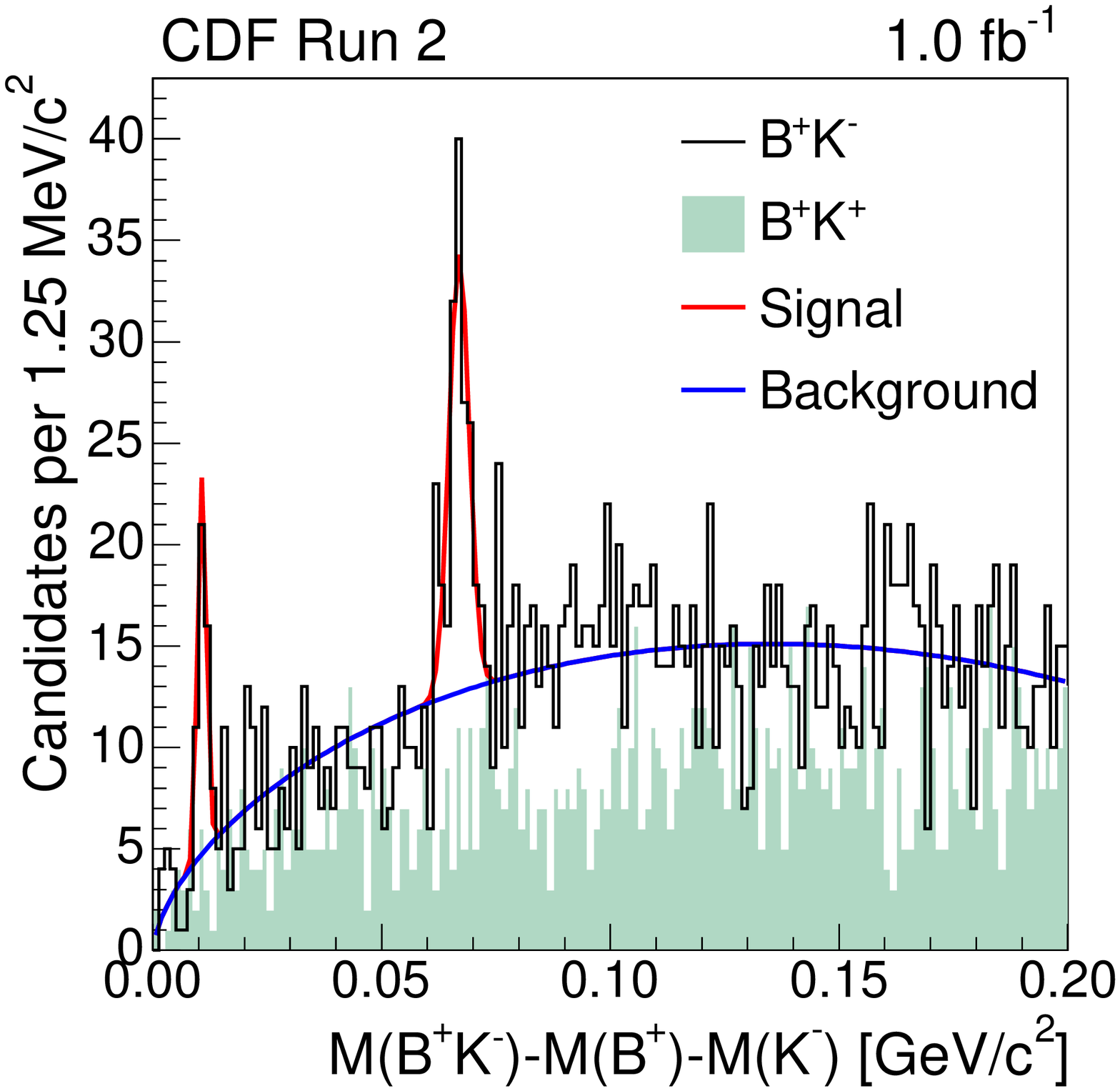}
\includegraphics[height=54mm]{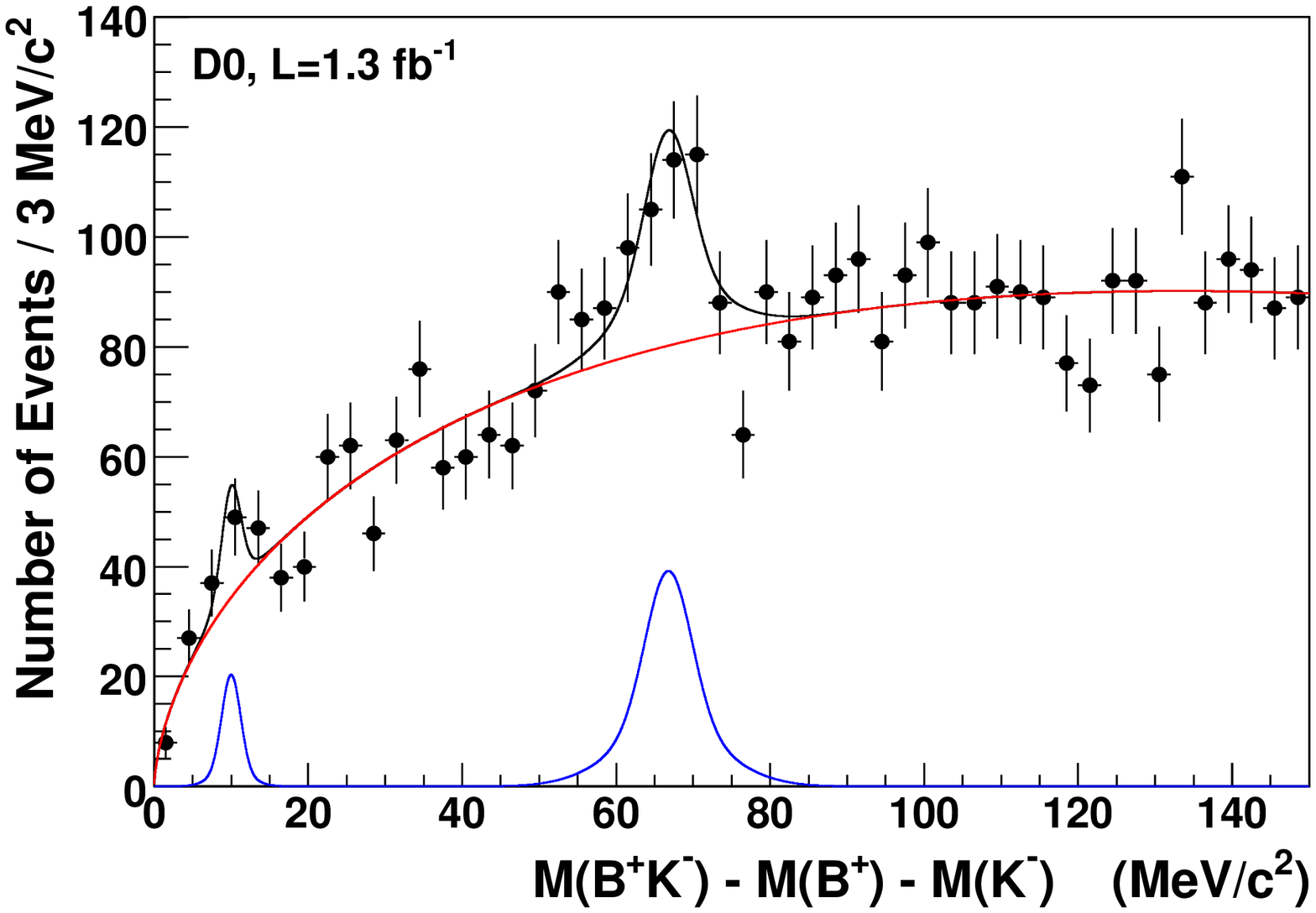}
\put(-360,130){\large\bf (a)}
\put(-185,120){\large\bf (b)}
\caption{Result of the fit to the $B_{s}^{**}$ mass difference $Q =
        m(BK)-m(B)-m(K)$ from (a) CDF and (b) D0.}
\label{fig:Bs_double_star}
\end{figure*}

$B_{s1}^0$ and $B_{s2}^{*0}$ candidates are reconstructed in the
following decay modes: $B_{s1}^0 \ra B^{*+}K^-$ with $B^{*+}\ra
B^+\gamma$ and $B_{s2}^{*0} \ra B^{*+}K^-$ with $B^{*+}\ra B^+\gamma$ as
well as $B_{s2}^{*0} \ra B^{+}K^-$. In both cases the soft photon from
the $B^*$ decay is not reconstructed resulting in a shift in the mass
spectrum.  D0 selects $B^+$ candidates in the fully reconstructed mode
$B^+\ra J/\psi K^+$ with $J/\psi\ra\mu^+\mu^-$, while CDF reconstructs
$B^+$ mesons in addition through the $B^+\ra \bar{D}^0\pi^+$ mode with
$\bar{D}^0\ra K^-\pi^+$. The CDF and D0 measurements are based on 1.0
and 1.3~fb$^{-1}$ of Run\,II data, respectively. The CDF
analysis~\cite{ref:CDF_Bss} finds $\sim\!31\,000$~$B^+\ra J/\psi K^+$
events and $\sim\!27\,200$ candidates in the $B^+\ra \bar{D}^0\pi^+$
channel. The D0 measurement~\cite{ref:D0_Bss} uses a signal of
$20\,915\pm293\pm200$ $B^+$ events from the decay $B^+\ra J/\psi K^+$.
Both experiments present their mass distributions in the
quantity $Q = m(BK)-m(B)-m(K)$ as displayed in
Figure~\ref{fig:Bs_double_star}(a) and (b).

A clear signal at $Q\sim67$~\mevcc\ is observed by CDF and D0 (see
Fig.~\ref{fig:Bs_double_star}), which is interpreted as the
$B_{s2}^{*0}$ state.  CDF reconstructs $95\pm23$ events in the peak at
$Q=(67.0\pm0.4\pm0.1)~\mevcc$ while D0 reports $125\pm25\pm10$ events at
$Q=(66.7\pm1.1\pm0.7)~\mevcc$.  In addition, CDF observes $36\pm9$
events in a peak at $Q=(10.7\pm0.2\pm0.1)~\mevcc$ which is the first
observation of this state interpreted as $B_{s1}^0$. A similar structure
in the Q~value distribution from D0 has a statistical significance of
less than $3\,\sigma$.  The measured masses are reported as
$m(B_{s2}^{*0})=(5839.6\pm1.1\pm0.7)~\mevcc$ from D0, while CDF quotes
$m(B_{s1}^0)=(5829.4\pm0.7)~\mevcc$ and
$m(B_{s2}^{*0})=(5839.6\pm0.7)~\mevcc$, where the statistical and
systematic errors are added in quadrature. The results from CDF and D0
are in good agreement.

\subsection{\boldmath{$\Bc$}~Meson Properties}

The \Bc~meson with a quark content $|b\bar c\,\rangle$ is a unique
particle as it contains two heavy quarks that can each decay via the
weak interaction. This means transitions of the $b$ or $c$ quark
contribute to the decay width of this meson. The \Bc~decay can occur via
the $b$~quark in a $b\ra c$~transition with a $J/\psi$ in the final
state (hadronic $J/\psi X$ or semileptonic $J/\psi\ell\nu X$ which is
the mode in which the \Bc~meson was discovered by CDF in
Run\,I~\cite{Abe:1998wi}) or via the
$\bar c$~quark in a $\bar c\ra \bar s$~transition with a $\bar B_s^0$ in
the final state (hadronic $\bar B_s^0 X$ or semileptonic $\bar
B_s^0\ell\nu X$). In addition, the $b\bar c$ quark pair can annihilate
into a $W$~boson with a lepton or quark pair coupling to the $W$ for a
$\Bc\ra\ell^-\bar\nu_{\ell}$ or $\Bc\ra q\bar q X$ transition. The
decays of both heavy quarks suggest copious decay modes and an expected
lifetime much shorter than that of other $B$~mesons. The lifetime of the
\Bc~meson is thus predicted from theory to be around
0.5~ps~\cite{Kiselev:2000pp}, while a measurement of the \Bc~mass tests
potential model predictions as well as lattice QCD calculations.

\subsubsection{Mass of the \Bc~Meson} 

The mass of the \Bc~meson has been predicted using a variety of
theoretical techniques. Non-relativistic potential
models~\cite{ref:Bc_potential} have been used to predict a mass of the
\Bc\ in the range 6247-6286~\mevcc, and a slightly higher value is found
for a perturbative QCD calculation~\cite{ref:Bc_QCD}. Recent lattice QCD
determinations provide a \Bc~mass prediction of
$(6304\pm12^{+18}_{-0})~\mevcc$~\cite{ref:Bc_lattice}. Precision
measurements of the properties of the \Bc~meson are thus needed to test
these predictions.

\begin{figure*}[tb]
\centering
\includegraphics[height=44mm]{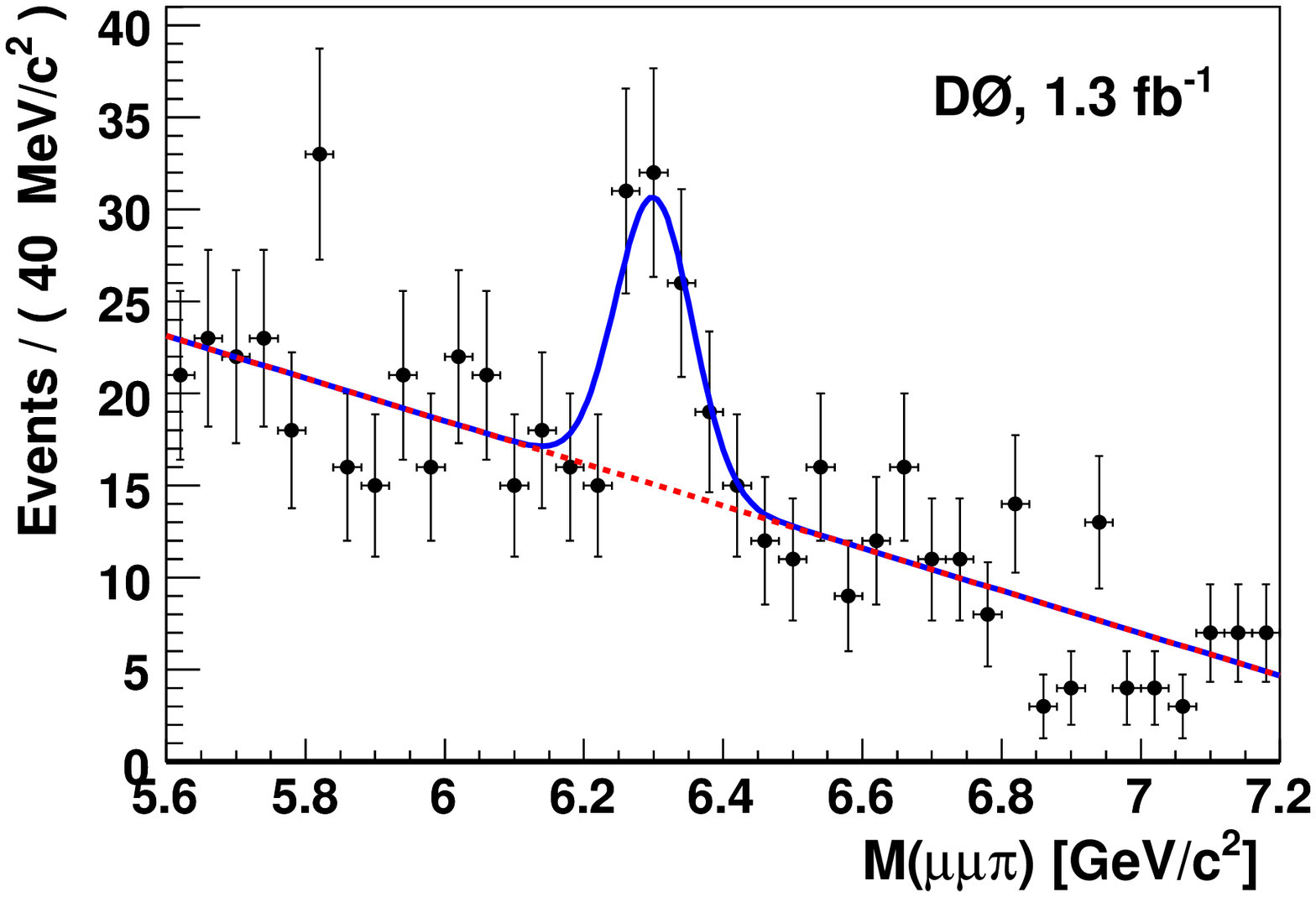}\ \
\includegraphics[height=47mm]{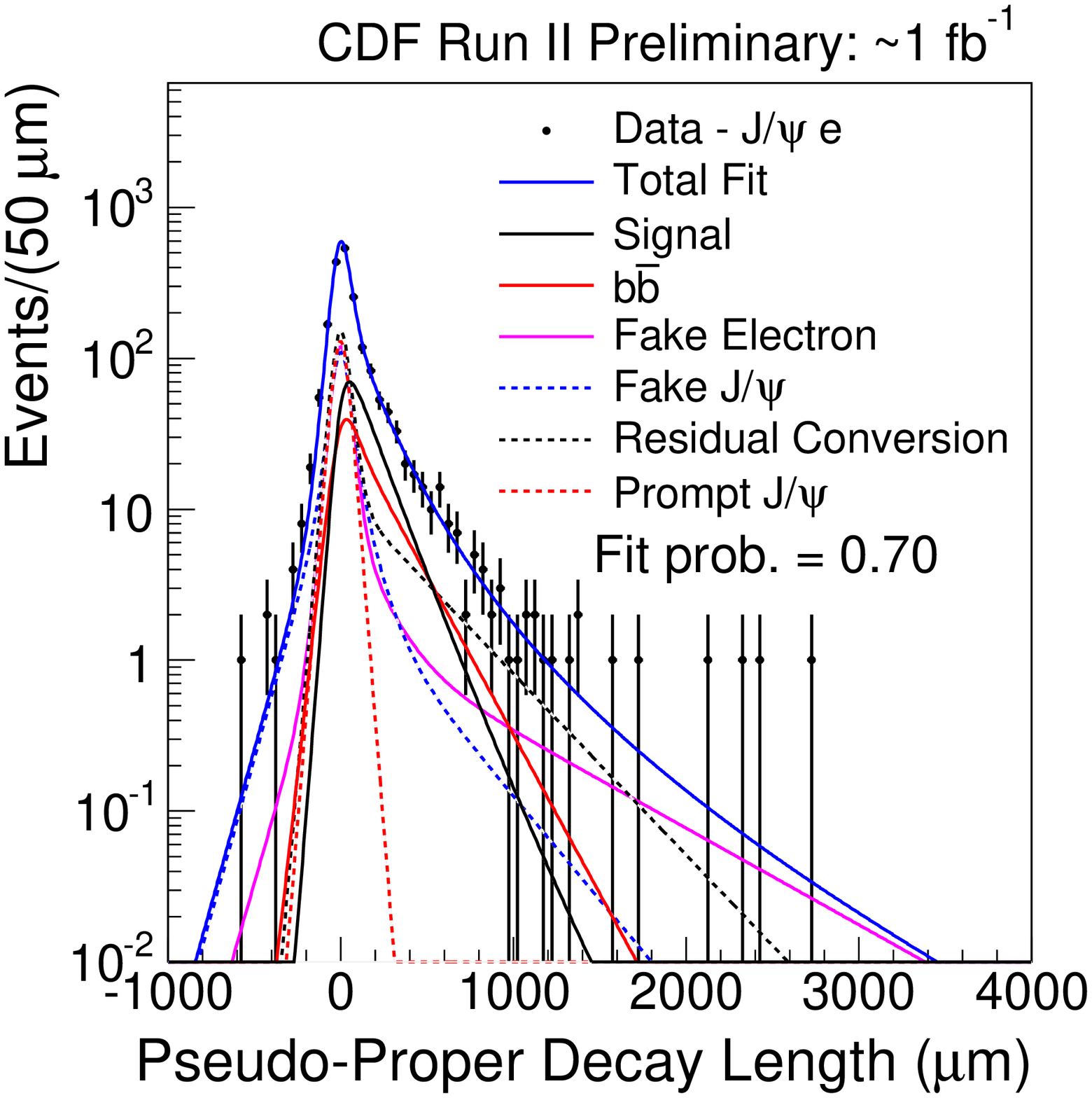}\ \ \ \ \ \
\includegraphics[height=47mm]{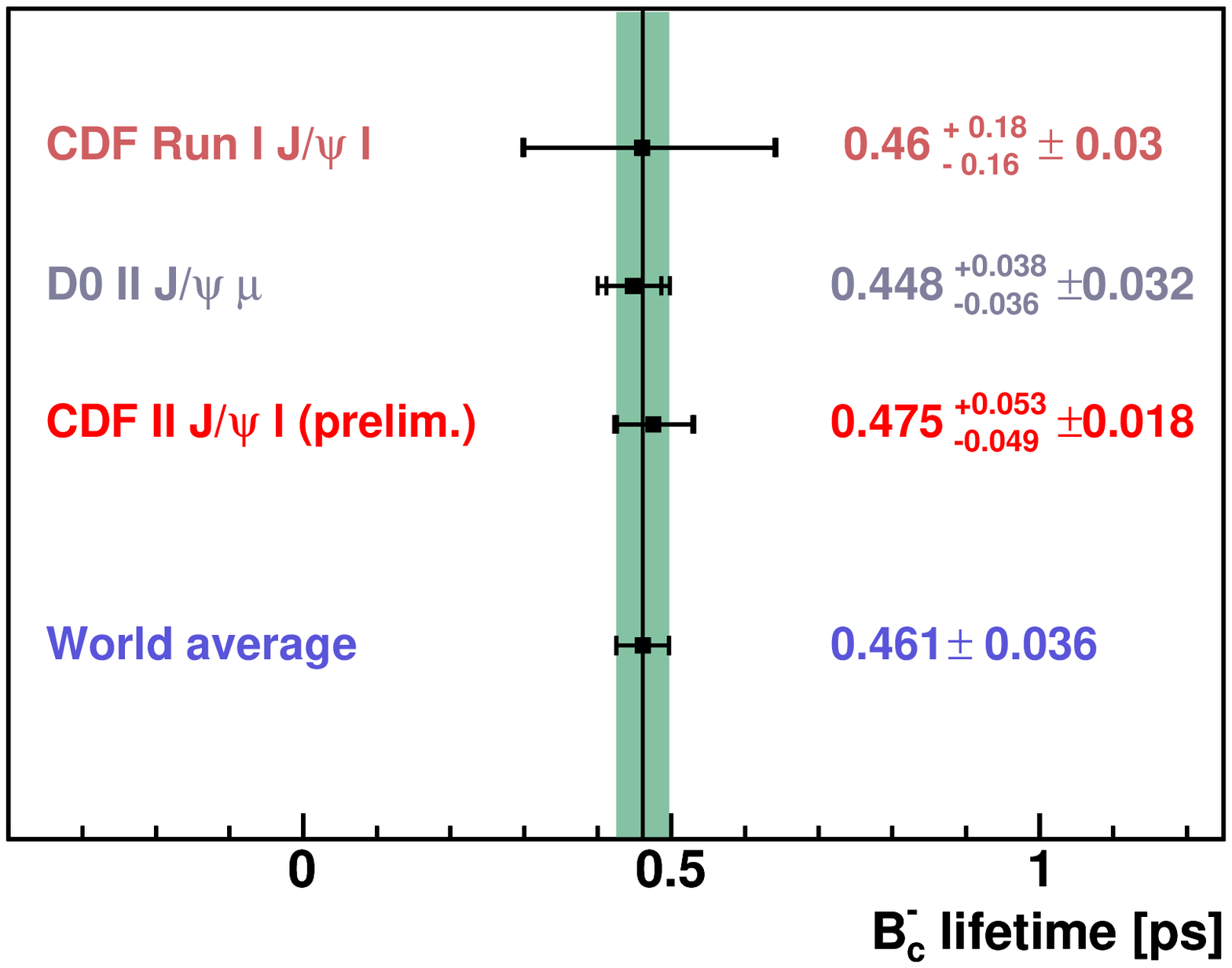}
\put(-360,90){\large\bf (a)}
\put(-220,105){\large\bf (b)}
\put(-163,27){\large\bf (c)}
\caption{(a) $J/\psi\,\pi^-$~invariant mass distribution of
  \Bc~candidates from D0. (b) Lifetime distribution from $\Bc\ra J/\psi
  e^-X$ with fit results overlaid from CDF. (c) Compilation of
  \Bc~lifetime measurements.}
\label{fig:Bc}
\end{figure*}

CDF and D0 both use fully reconstructed $\Bc\ra
J/\psi\,(\ra\mu^+\mu^-)\,\pi^-$ decays for a precise measurement of the
\Bc~mass.  CDF first published their analysis~\cite{ref:CDF_Bcmass},
based on 2.4~fb$^{-1}$ of data, where the \Bc~selection is optimized on
the signal yield of $B^-\ra J/\psi K^-$ and the obtained selection
criteria are directly transferred to the $J/\psi \pi^-$ data for an
unbiased selection.  A signal of $108\pm15$ events with a significance
greater than $8\,\sigma$ is observed. The mass of the \Bc~meson is
measured to be $(6275.6\pm2.9\pm2.5)~\mevcc$.  To test the background
reduction process, the D0 analysis~\cite{ref:D0_Bcmass}, based on
1.3~fb$^{-1}$ of data, uses a well-understood signal sample of $B^-\ra
J/\psi K^-$ data. After the final selection the
$J/\psi\,\pi^-$~invariant mass distribution of \Bc~candidates from D0 is
obtained as shown in Figure~\ref{fig:Bc}(a). An unbinned likelihood fit
yields a signal of $54\pm12$~events corresponding to a significance of
$5.2\,\sigma$. The extracted \Bc~mass value is reported as
$(6300\pm14\pm5)~\mevcc$. Combining both results yields a world average
\Bc~mass of $m(\Bc)=(6276\pm4)~\mevcc$.  In comparison to 
theoretical
predictions~\cite{ref:Bc_potential,ref:Bc_QCD,ref:Bc_lattice},  
the experimental measurements, especially the CDF result
with small uncertainties, start to challenge the predictions of
theoretical models and lattice QCD calculations.

\subsubsection{Lifetime of the \Bc~Meson} 

As discussed above, the decay of a \Bc~meson can occur via a $b\ra
c$~transition, a $\bar c\ra \bar s$~transition or a $b\bar c$ quark pair
annihilation into a $W$~boson leading to an expected lifetime of order
$(0.5\pm0.1)$~ps~\cite{Kiselev:2000pp}, only one third that of other
$B$~mesons.  Due to the large branching fraction, CDF and D0 both use
the semileptonic decay $\Bc\ra J/\psi\ell^-\nu X$ with
$J/\psi\ra\mu^+\mu^-$ for their measurement of the \Bc~lifetime. The
main issue in using a \Bc~semileptonic decay is to control the
backgrounds since no \Bc~mass peak can be reconstructed. Since the event
signature is two muons forming the $J/\psi$~meson plus a third lepton,
the background sources are fake $J/\psi$'s, fake leptons, or
uncorrelated real $J/\psi$'s and leptons from $b\bar b$~events where the
$J/\psi$ is from one $b$~quark while the lepton is from the other
$b$~quark jet. In the $J/\psi e^-$ channel, which is only used by CDF,
there is an additional background of electrons from residual photon
conversions $\gamma\ra e^+e^-$ within the detector material. CDF and D0
estimate the backgrounds with Monte Carlo or based on data depending on
the analysis approach. The D0 analysis~\cite{abazov:2008rba}, using
1.3~fb$^{-1}$ of data, identifies $881\pm80$ $J/\psi\mu X$ signal
candidates and measures the \Bc~lifetime as
$\tau(\Bc)=(0.448^{+0.038}_{-0.036}\pm0.032)$~ps.  

In a preliminary analysis, based on 1~fb$^{-1}$ of data, CDF analyzes
both the $J/\psi\mu$ and $J/\psi e$ final state in the $J/\psi\ell$
invariant mass range between 4-6~\gevcc, in which the \Bc~signal is
expected to lie. The CDF lifetime distribution with the various
background sources indicated is shown in Figure~\ref{fig:Bc}(b) for the
$J/\psi e$~mode. Combining the electron and muon channel, CDF measures
$\tau(\Bc)=(0.475^{+0.053}_{-0.049}\pm0.018)$~ps. A compilation of both
measurements together with an older CDF Run\,I result is displayed in
Figure~\ref{fig:Bc}(c) and a world average \Bc~lifetime of
$\tau(\Bc)=(0.461\pm0.036)$~ps is determined in good agreement with
theoretical predictions.

\section{PROPERTIES OF BOTTOM BARYONS} 

The QCD treatment of quark-quark interactions significantly simplifies
if one of the participating quarks is much heavier than the QCD
confinement scale $\Lambda_{\rm QCD}$.  In the limit of
${m_Q\ra\infty}$, where ${m_Q}$ is the mass of the heavy quark, the
angular momentum and flavour of the light quark become good quantum
numbers. This approach, known as heavy quark effective theory, thus
views a baryon made out of one heavy quark and two light quarks as
consisting of a heavy static color field surrounded by a cloud
corresponding to the light di-quark system. The two quarks form either a
$\bar{3}$ or $6$ di-quark under SU(3), according to the decomposition 
$3\otimes 3 = \bar{3} \oplus 6$, leading to a generic scheme of baryon
classification. Di-quark states containing quarks in an antisymmetric
flavour configuration, $[q_1,q_2]$, are called $\Lambda$-type whereas
states with di-quarks containing quarks in a flavour symmetric state,
$\{q_1,q_2\}$, are called $\Sigma$-type. For baryons with a bottom
quark, this classification gives the ground state \Lb~baryon with quark
content $|bdu\,\rangle$ and the \Sb~baryons with quark content
$\Sb^{(*)+}=|buu\,\rangle$ and $\Sb^{(*)-}=|bdd\,\rangle$.  If one of
the light quarks is a strange quark, we classify the bottom baryon as a
cascade $\Xi_b$~baryon and the double strange bottom baryon is the
$\Omb$ with quark content $|bss\,\rangle$.

\subsection{The \boldmath{\Lb} Lifetime Story} 

\begin{figure*}[tb]
\centering
\includegraphics[height=65mm]{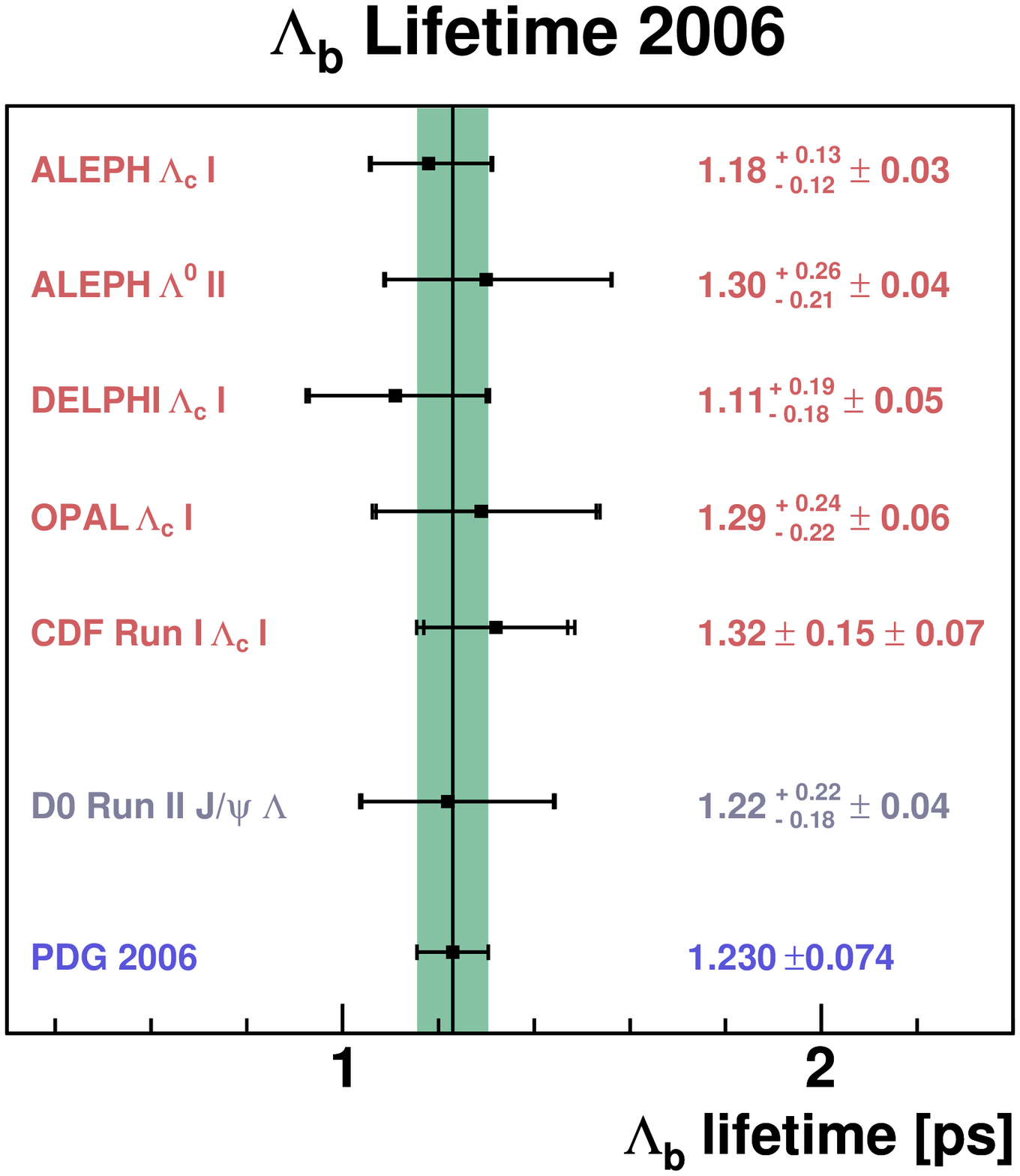}
\includegraphics[height=63mm]{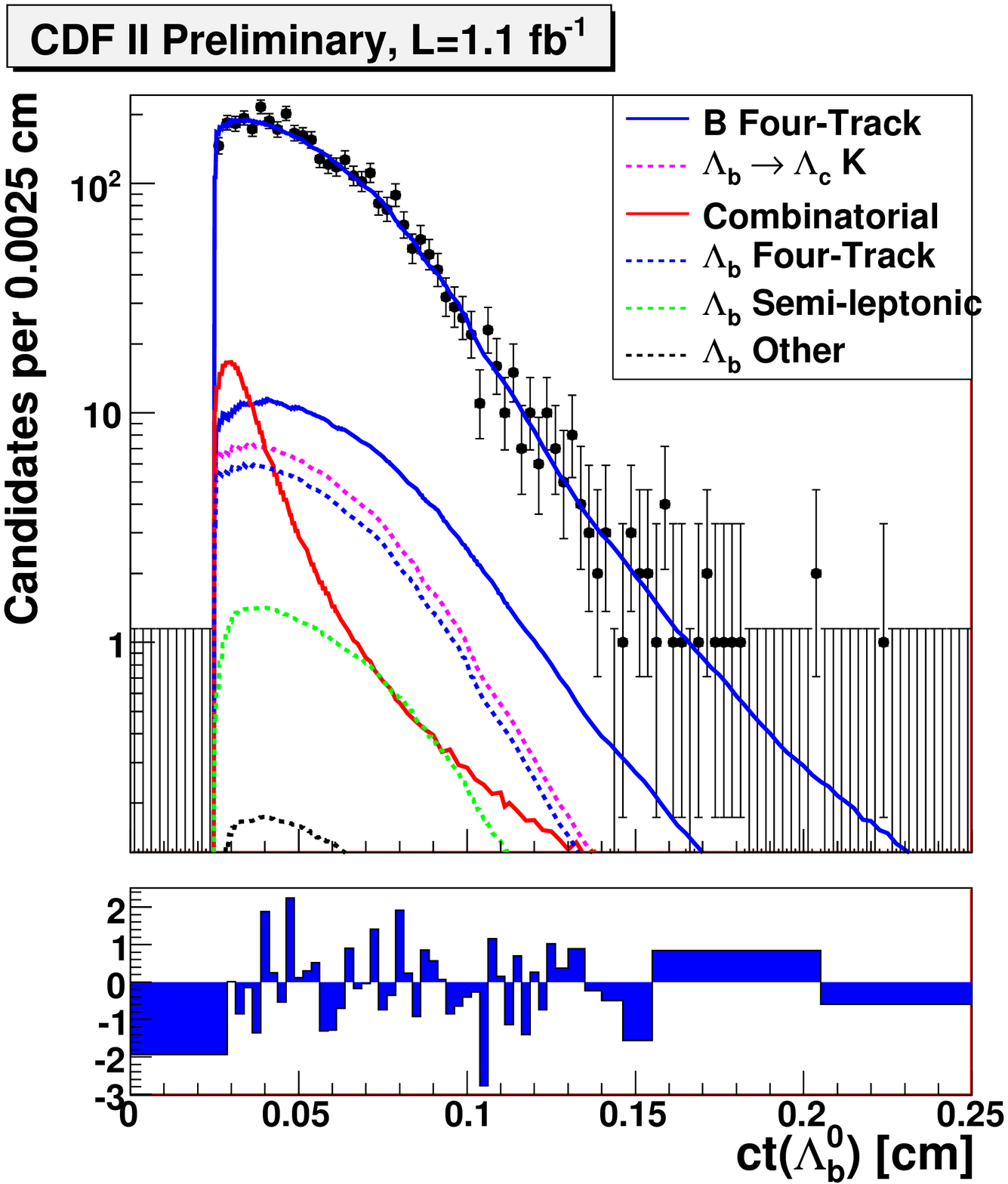}\ \
\ \ \
\includegraphics[height=65mm]{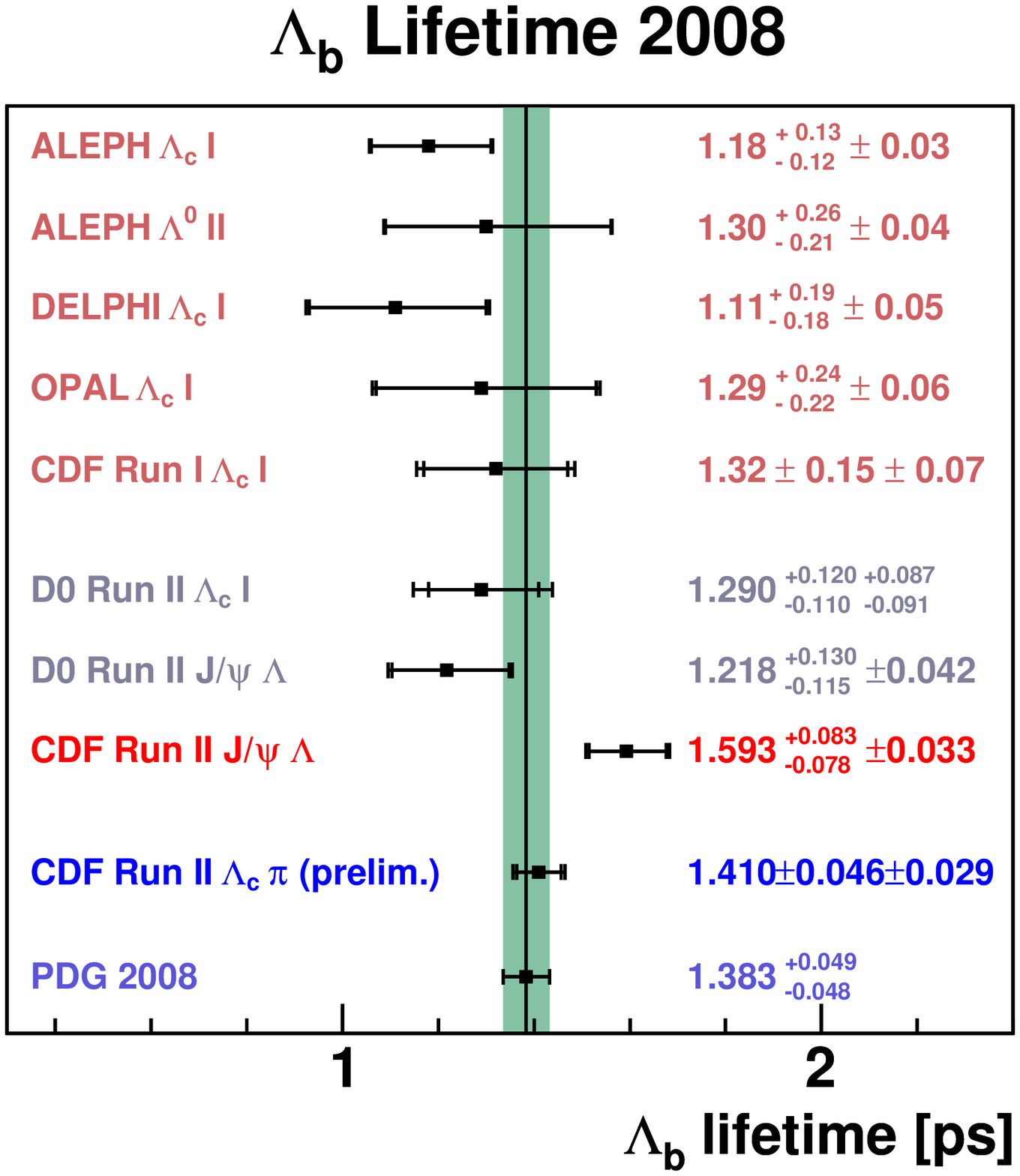}
\put(-410,152){\large\bf (a)}
\put(-265,150){\large\bf (b)}
\put(-122,152){\large\bf (c)}
\caption{(a) Summary of \Lb~lifetime measurements as of 2006. (b)
  \Lb~lifetime fit of CDF data using $\Lb\ra\Lambda_c^+\pi^-$ and (c)
  compilation of \Lb~lifetime measurements as of 2008.}
\label{fig:Lb}
\end{figure*}

The mass of the ground state bottom baryon, the
\Lb~$(\,|bdu\,\rangle\,)$, has been established for quite some time with
the current mass value
$m(\Lb)=(5620.2\pm1.6)~\mevcc$~\cite{ref:PDG2008}. However, the
lifetime of the \Lb~baryon has been puzzling the community for a long
time. The situation of \Lb~lifetime measurements as of 2006 is
summarized in Figure~\ref{fig:Lb}(a). A world average lifetime of
$\tau(\Lb)=(1.230\pm0.074)$~ps is quoted by the PDG in 2006 based on
several LEP measurements, one CDF Run\,I measurement and the first
Run\,II measurement with 0.25~fb$^{-1}$ from D0~\cite{Abazov:2004bn}
using the decay mode $\Lb\ra J/\psi\Lambda^0$. The 2006 world average
\Lb~lifetime translates into a lifetime ratio
$\tau(\Lb)/\tau(\Bz)=0.804\pm0.049$. This number is on the low side of
theoretical predictions which are in the range of $0.88\pm0.05$.  This
introduces the long-standing puzzle that the \Lb~lifetime is measured
smaller than theoretical predictions.

Then in 2007, the D0 experiment updated its measurement of the
\Lb~lifetime from $J/\psi\Lambda^0$ with 1.2~fb$^{-1}$ of
data~\cite{Abazov:2007sf} measuring
$\tau(\Lb)=(1.218^{+0.130}_{-0.115}\pm0.042)$~ps resulting in a lifetime
ratio $\tau(\Lb)/\tau(\Bz)=0.811^{+0.096}_{-0.087}\pm0.034$. In the same
year D0 released another measurement~\cite{abazov:2007th} of the
\Lb~lifetime using semileptonic decays $\Lb\ra\mu^-\bar\nu\Lambda_c^+X$.
In 1.2~fb$^{-1}$ of $p\bar p$~collision data, D0 identifies $4437\pm329$
signal candidates and measures
$\tau(\Lb)=(1.290{^{+0.119}_{-0.111}}{^{+0.087}_{-0.091}})$~ps. Both D0
results are in good agreement with the 2006 world average \Lb~lifetime.
In the same year in 2007, CDF published a measurement of the
\Lb~lifetime in the exclusive decay $\Lb\ra J/\psi\Lambda^0$ using
1.2~fb$^{-1}$ of data~\cite{Abulencia:2006dr}. The resulting
$\tau(\Lb)=(1.593^{+0.083}_{-0.078}\pm0.033)$~ps was the single most
precise measurement of $\tau(\Lb)$ but is $3.2\,\sigma$ higher than the
2006 world average. This surprising fact is also evident when forming a
ratio with the world average \Bz~lifetime yielding
$\tau(\Lb)/\tau(\Bz)=1.041\pm0.057$ larger than one! This result was a
big surprise and further measurements were needed to resolve the
situation.

New in 2008 is a preliminary CDF measurement of the \Lb~lifetime using
fully reconstructed $\Lb\ra\Lambda_c^+\pi^-$ decays with $\Lambda_c^+\ra
p K^+\pi^-$. With a dataset of 1.2~fb$^{-1}$, CDF obtains a clean sample
of about 3000 fully reconstructed \Lb~signal events. From the lifetime
distribution shown in Figure~\ref{fig:Lb}(b), CDF measures
$\tau(\Lb)=(1.410\pm0.046\pm0.029)$~ps and reports a lifetime ratio
$\tau(\Lb)/\tau(\Bz)=0.922\pm0.039$ in good agreement with theoretical
predictions. This measurement is as precise as and in good agreement
with the current world average of the \Lb~lifetime
$\tau(\Lb)=(1.383^{+0.049}_{-0.048})$~ps determined for the PDG 2008
edition~\cite{ref:PDG2008} without including the new preliminary CDF
result. The situation of \Lb~lifetime measurements is summarized in
Figure~\ref{fig:Lb}(c) where it can be seen that the new CDF result is
also in agreement within one standard deviation with the 2007
measurement of $\tau(\Lb)$ from CDF.  It appears that the longstanding
puzzle surrounding the \Lb~lifetime has been resolved.

\subsection{\boldmath{\Sb} and \boldmath{$\Sb^*$} Baryons}

Until recently only one bottom baryon, the \Lb, has been directly
observed.  The $\Sb^{(*)}$~baryon has quark content
$\Sb^{(*)+}=|buu\,\rangle$ and $\Sb^{(*)-}=|bdd\,\rangle$.  In the
$\Sigma$-type ground state, the light di-quark system has isospin $I=1$
and $J^P=1^+$.  Together with the heavy quark, this leads to a doublet
of baryons with $J^P=\frac{1}{2}^+$ (\Sb) and $J^P=\frac{3}{2}^+$
($\Sb^*$).  The ground state $\Sigma$-type baryons decay strongly to
$\Lambda$-type baryons by emitting pions.  In the limit
${m_Q\ra\infty}$, the spin doublet $\{\Sb,\Sb^*\}$ would be exactly
degenerate since an infinitely heavy quark does not have a spin
interaction with a light di-quark system.  As the heavy quark is not
infinitely massive, there will be a small mass splitting between the
doublet states resulting in an additional isospin splitting between the
$\Sb^{(*)-}$ and $\Sb^{(*)+}$ states~\cite{Rosner:1998zc}.  There
exist a number of predictions for the masses and isospin splittings of
these states using HQET, non-relativistic and relativistic potential
models, $1/{\rm N}_c$ expansion, sum rules and lattice QCD
calculations~\cite{{Rosner:1998zc},ref:Stanley:1980fe}.

\begin{figure*}[tb]
\centering
\includegraphics[height=73mm]{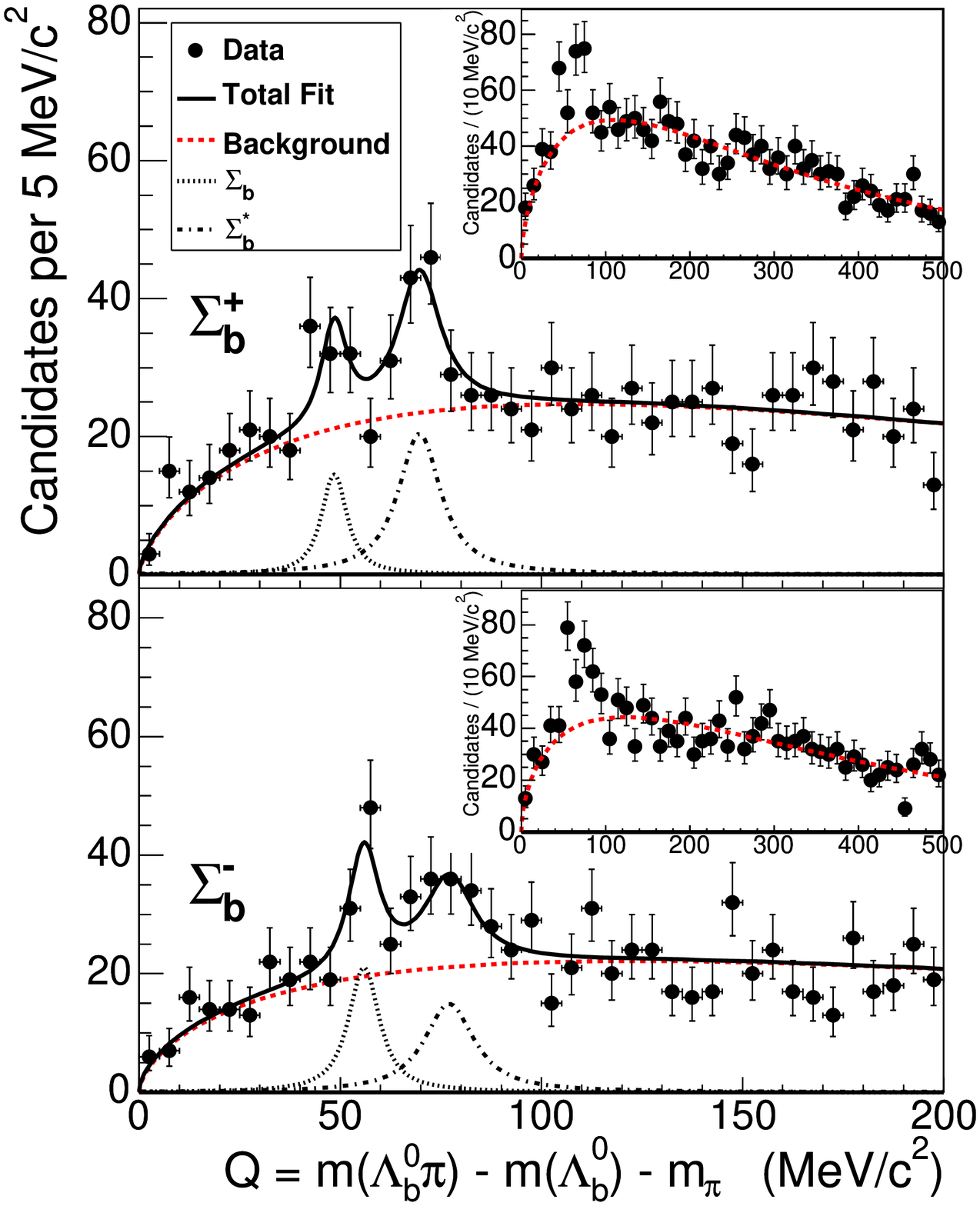}\quad
\includegraphics[height=68mm]{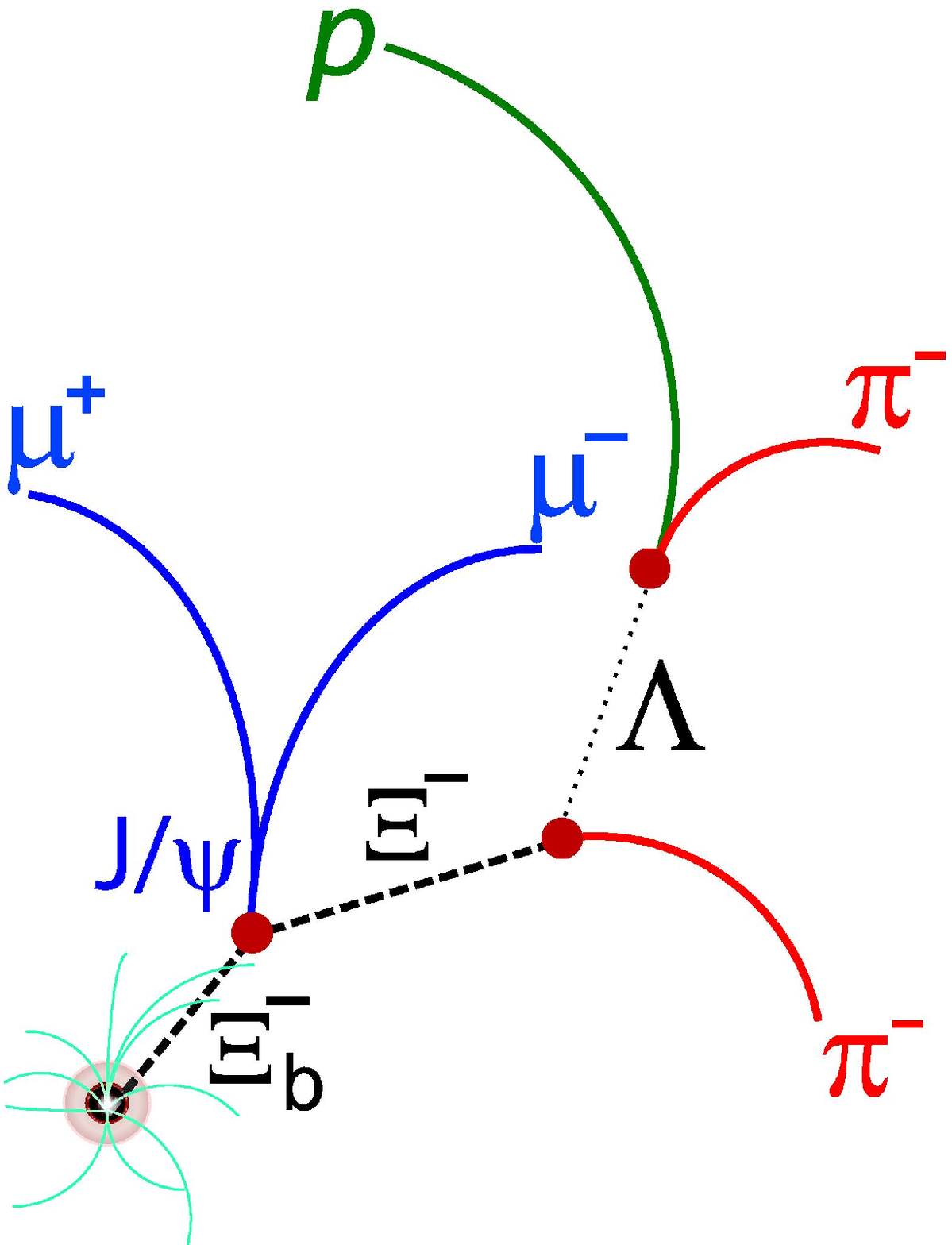}\quad
\includegraphics[height=73mm]{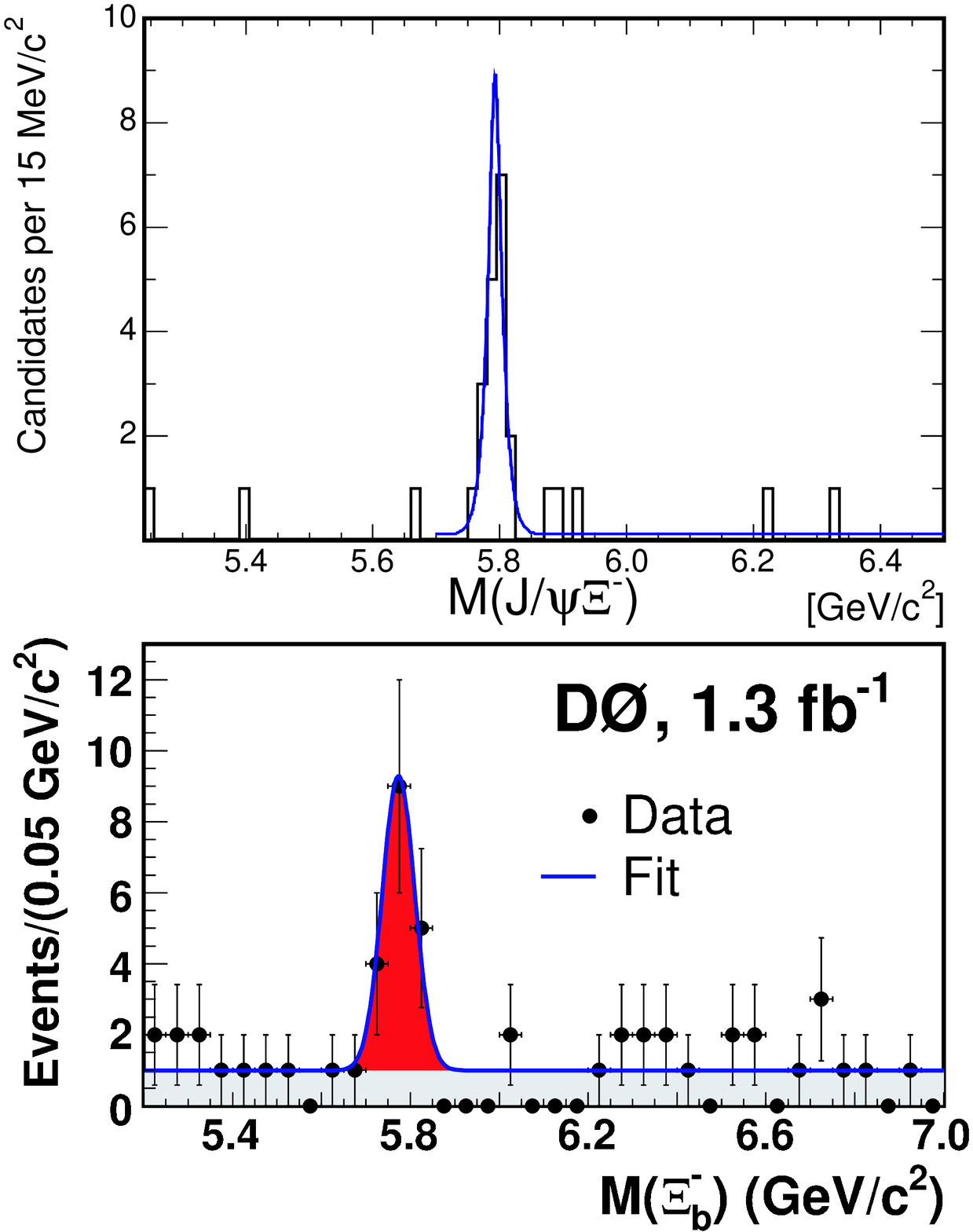}
\put(-360,192){\large\bf (a)}
\put(-315,192){\large\bf (b)}
\put(-133,188){\large\bf (c)}
\caption{(a) The $\Sb^{(*)}$ fit to the $\Lb\pi^+$~and
  $\Lb\pi^-$~subsamples. The top plot shows the $\Lb\pi^+$~data,
  which contain $\Sb^{(*)+}$, while the bottom plot shows the
  $\Lb\pi^-$~subsample, which contains $\Sb^{(*)-}$.  (b) Schematic of
  the $\Xib\ra J/\psi\,\Xi^-$ decay topology.  (c)~The $J/\psi\,\Xi^-$
  invariant mass distribution from CDF (top) and D0 (bottom) including
  fits to the data overlaid.}
\label{fig:Sigmab_Xib}
\end{figure*}

The CDF collaboration has accumulated a large data sample of \Lb~baryons
using the CDF displaced track trigger.  Using a 1.1~fb$^{-1}$ data set
of fully reconstructed $\Lb\ra\Lambda_c^+\pi^-$ candidates, CDF searches
for the decay $\Sb^{(*)\pm} \ra \Lb\pi^{\pm}$. The CDF
analysis~\cite{ref:CDF_sigmab} reconstructs a \Lb~yield of approximately
2800 candidates in the signal region $m(\Lb)\in [5.565, 5.670]~\gevcc$.
To separate out the resolution on the mass of each \Lb~candidate, CDF
searches for narrow resonances in the mass difference distribution of $Q
= m(\Lb \pi) - m(\Lb) - m(\pi)$.  Unless explicitly stated,
$\Sb^{(*)}$~refers to both the $J=\frac{1}{2}$ ($\Sb^{\pm}$) and
$J=\frac{3}{2}$ ($\Sb^{*\pm}$) states while the analysis distinguishes
between $\Sb^{(*)+}$ and $\Sb^{(*)-}$.  There is no transverse momentum
cut applied to the pion from the $\Sb^{(*)}$~decay, since these tracks
are expected to be very soft.  The result of the $\Sb^{(*)}$ search in
the $\Lb\pi^+$~and $\Lb\pi^-$~subsamples is displayed in
Figure~\ref{fig:Sigmab_Xib}(a).  The top plot shows the
$\Lb\pi^+$~subsample, which contains $\Sb^{(*)+}$, while the bottom plot
shows the $\Lb\pi^-$~subsample, which contains $\Sb^{(*)-}$.  The insets
show the expected background plotted on the data for $Q \in$ [0, 500]
$\mevcc$, while the signal fit is shown on a reduced range of $Q \in$
[0, 200]~\mevcc.  The final fit results for the $\Sb^{(*)}$~measurement
are summarized in Table~\ref{tab:sigmab}. The absolute $\Sb^{(*)}$~mass
values are calculated using a CDF measurement of the
\Lb~mass~\cite{ref:Acosta:2005mq}, which contributes to the systematic
uncertainty. The mass splitting $\Delta_{\Sb^*}$ between $\Sb^*$ and
\Sb\ has been set in the fit to be the same for $\Sb^+$ and $\Sb^-$.

\begin{table}[tb]
\begin{center}
  \caption{Final results for the $\Sb^{(*)}$~mass measurements.  The
    first uncertainty is statistical and the second is systematic.  The
    absolute $\Sb^{(*)}$~mass values are calculated using a CDF measurement of
    the \Lb~mass~\cite{ref:Acosta:2005mq}.}
\begin{tabular}{l l l l} 
\hline
State           & \multicolumn{1}{c}{Yield} 
& \multicolumn{1}{c}{$Q$ or $\Delta_{\Sb^*}$ [$\mevcc$]} & \multicolumn{1}{c}{Mass [$\mevcc$]}  \\
\hline
${\Sb^+}$       & $32^{+13+5}_{-12-3}$  & $Q_{\Sb^+} = 48.5^{+2.0+0.2}_{-2.2-0.3}$      & $5807.8^{+2.0}_{-2.2}\pm 1.7$ \\
${\Sb^-}$       & $59^{+15+9}_{-14-4}$  & $Q_{\Sb^-} = 55.9\pm 1.0\pm 0.2$              & $5815.2\pm 1.0\pm 1.7$      \\
${\Sb^{*+}}$    & $77^{+17+10}_{-16-6}$ & $\Delta_{\Sb^*} = 21.2^{+2.0+0.4}_{-1.9-0.3}$ & $5829.0^{+1.6+1.7}_{-1.8-1.8}$ \\
${\Sb^{*-}}$    & $69^{+18+16}_{-17-5}$ &       & $5836.4\pm 2.0^{+1.8}_{-1.7}$ \\
\hline
\end{tabular}
\label{tab:sigmab}
\end{center}
\end{table}

\subsection{Observation of  the \boldmath{\Xib} Baryon}

The $\Xi_b$~baryons with a quark content of $\Xib=|bds\,\rangle$ and
$\Xi_b^0=|bus\,\rangle$ decay weakly through the decay of the $b$~quark
and are expected to have a lifetime similar to the typical $B$~hadron
lifetime of about 1.5~ps. Possible decay modes of the $\Xi_b^0$ include
$\Xi_b^0\ra\Xi_c^0\pi^0$ or $J/\psi\,\Xi^0\, (\ra\Lambda^0\pi^0)$. Both
decays involve the reconstruction of neutral pions which is difficult to
achieve at CDF and D0. However, the \Xib\ can decay through $\Xib\ra
J/\psi\Xi^-$ followed by $\Xi^-\ra\Lambda^0\pi^-$ with $\Lambda^0\ra
p\pi^-$ and $J/\psi\ra\mu^+\mu^-$. This is the decay mode in which CDF
and D0 search for the \Xib~baryon.

A schematics of the decay topology is shown in
Figure~\ref{fig:Sigmab_Xib}(b) from where the challenges in the
\Xib~reconstruction become apparent. The \Xib~baryon travels an average
distance of $c\tau(\Xib)\sim450~\mu$m and then decays into a $J/\psi$
and $\Xi^-$ which has a $c\tau(\Xi^-)=4.9$~cm traversing parts of the
silicon detector. Furthermore, the $\Xi^-$ decays into a $\Lambda^0$
which has a $c\tau(\Lambda^0)=7.9$~cm often decaying in the inner layers
of the main tracker. This brings significant challenges for the
reconstruction of the \Xib\ decay products and their track
reconstruction. The D0~analysis~\cite{ref:D0_Xib} based on 1.3~fb$^{-1}$
of data runs a special re-processing of the dimuon datasets to improve
the efficiency of reconstructing high impact parameter tracks in the
track pattern recognition algorithm. The event selection is based on
wrong-sign data and guided by \Xib~Monte Carlo events. On the other
hand, CDF develops a dedicated silicon-only tracking algorithm to
reconstruct the charged $\Xi^-$~tracks in its silicon tracker. The CDF
event selection~\cite{ref:CDF_Xib} based on 1.9~fb$^{-1}$ of data uses a
$B^-\ra J/\psi K^-$ control sample where the selection criteria are
developed. The $K^-$ is then replaced in the data analysis by the
$\Xi^-$ for an unbiased event selection.

Both experiments observe significant \Xib~signals as can be seen in the
$J/\psi\,\Xi^-$ invariant mass distribution in
Figure~\ref{fig:Sigmab_Xib}(c). D0 finds $15.2\pm4.4^{+1.9}_{-0.4}$
\Xib~signal event with a Gaussian significance of $5.2\,\sigma$ and
reports a mass of $m(\Xib)=(5774\pm11\pm15)~\mevcc$~\cite{ref:D0_Xib}.
CDF observes $17.5\pm4.3$ \Xib~signal event with a Gaussian significance
of $7.7\,\sigma$ and measures a \Xib~mass of
$m(\Xib)=(5792.9\pm2.5\pm1.7)~\mevcc$~\cite{ref:CDF_Xib}.  In addition,
D0 verifies that the lifetime of the \Xib~candidates is compatible with
a $B$~hadron like lifetime.

Soon after this conference, the D0 collaboration
announced the observation of another heavy bottom
baryon~\cite{Abazov:2008qm}, the double strange $\Omb$ baryon with quark
content $|bss\,\rangle$. With the same dataset as used for the
\Xib~observation based on 1.3~fb$^{-1}$ of $p\bar p$~collisions, D0
reconstructs $\Omb\ra J/\psi\Omega^-$ followed by $\Omega^-\ra\Lambda^0
K^-$ and obtains a mass measurement of
$m(\Omb)=(6165\pm10\pm13)~\mevcc$ based on an \Omb~signal of
$17.8\pm4.9\pm0.8$ events. The significance of the observed signal is
$5.4\,\sigma$ corresponding to a probability of $6.7\times10^{-8}$ of
it arising from background fluctuation.

\section{CONCLUSION}

We have reviewed recent result on heavy $B$~hadron properties focusing
on Run\,II measurements from the Fermilab Tevatron which offers a rich
heavy flavour program. A wealth of new results on properties of heavy
$B$~hadron states from CDF and D0 has been available. These include
measurements of the lifetime and decay width difference \dGs\ in
\Bs~meson decays, updates on $CP$~violation in \BsJPsiPhi~decays which
continue to show an intriguing discrepancy with the standard model
prediction. In addition, the Belle collaboration used \Bs~mesons
produced at the $\Upsilon(5S)$ resonance to obtain competitive branching
ratio measurements for \Bs~decays. We also reviewed recent results on
the mass and lifetime of the \Bc~meson. With respect to bottom baryons,
the puzzle of the \Lb~lifetime measurements being lower than theoretical
predictions appears to be solved. New heavy bottom baryons have been
established, the $\Sb^{(*)}$~states as well as the \Xib\ and \Omb~baryon. We
expect more results from the Tevatron which will accumulate more data
until the end of Run\,II currently scheduled to conclude in 2010.  With
the onset of the Large Hadron Collider in 2009, more exciting result on
heavy $B$~hadron properties are expected, especially from the LHCb
experiment.

\begin{acknowledgments}
  This paper is dedicated to the memory of Michael P.~Schmidt, Randy
  Pausch and all whose lives are lost to cancer. There is hope. Our
  personal hope includes Holcombe Grier and his outstanding team at DFCI
  as well as the staff and all the wonderful volunteers at Camp
  Sunshine.

  I would like to thank the organizers of this stimulating meeting,
  especially Joe Kroll, Nigel Lockyer and Stew Smith, for the
  opportunity to present these results as well as an outstanding
  conference organization. I am grateful to my colleagues from the
  \babar, Belle, CDF and D0~collaborations for their help in preparing
  this talk.  This work was supported in part by the U.S.~Department of
  Energy under Grant No.~DE-FG02-91ER40682.
\end{acknowledgments}


\end{document}